\documentclass[aps,10pt,prd,twocolumn,superscriptaddress,amssymb,amsmath,nofootinbib]{revtex4-2}

\usepackage[pdftex,breaklinks,colorlinks,
linkcolor=Blue,
citecolor=teal,
anchorcolor=red,
urlcolor=cyan,
pdfencoding=auto,
pdftitle={Sourced metric perturbations of Kerr spacetime in Lorenz gauge},
pdfauthor={Barry Wardell, Chris Kavanagh, Sam R. Dolan},
pdfdisplaydoctitle]{hyperref}
\usepackage[usenames,dvipsnames,svgnames,table]{xcolor}
\usepackage{mathrsfs}
\usepackage{multirow}
\usepackage{graphicx}
\usepackage{IEEEtrantools}

\usepackage{orcidlink}

\font\ec=ecrm0800 at 11pt
\def\th{\hbox{\ec\char'336}}
\def\edth{\hbox{\ec\char'360}}

\newcommand{\mb}{{\bar{m}}}
\newcommand{\Lor}{{\rm L}}
\newcommand{\LieT}{\pounds_{\mathrm{T}}}
\newcommand{\AAB}{{\rm AAB}}
\newcommand{\trace}{{\rm Trace}}
\newcommand{\TF}{{\rm TF}}
\newcommand{\maxwell}{{\rm TF}}
\newcommand{\Weyl}{{\cC}}
\newcommand{\DKW}{{\rm DKW}}
\newcommand{\SE}{{\rm T}}
\newcommand{\TKV}{\mathrm{T}}
\newcommand{\DKWSE}{{\DKW,\SE}}
\newcommand{\trev}[1]{\hat{#1}}

\newcommand{\cA}{\mathcal{A}}
\newcommand{\cB}{\mathcal{B}}
\newcommand{\cC}{\mathcal{C}}
\newcommand{\cF}{\mathcal{F}}
\newcommand{\cH}{\mathcal{H}}
\newcommand{\cJ}{\mathcal{J}}
\newcommand{\cW}{\mathcal{W}}
\newcommand{\cZ}{\mathcal{Z}}

\newcommand{\cS}{\mathcal{S}}
\newcommand{\cE}{\mathcal{E}}
\newcommand{\cO}{\mathcal{O}}
\newcommand{\cT}{\mathcal{T}}
\newcommand{\cN}{\mathcal{N}}
\newcommand{\cK}{\mathcal{K}}
\newcommand{\cP}{\mathcal{P}}
\newcommand{\cD}{\mathcal{D}}

\newcommand{\Curl}{\mathscr{C}}
\newcommand{\CurlDag}{\mathscr{C}^\dag}
\newcommand{\Div}{\mathscr{D}}
\newcommand{\Twist}{\mathscr{T}}

\begin{document}

\title{Sourced metric perturbations of Kerr spacetime in Lorenz gauge}

\author{Barry Wardell\,\orcidlink{0000-0001-6176-9006}}
\affiliation{School of Mathematics and Statistics, University College Dublin, Belfield, Dublin 4, Ireland.}

\author{Chris Kavanagh\,\orcidlink{0000-0002-2874-9780}}
\affiliation{School of Mathematics and Statistics, University College Dublin, Belfield, Dublin 4, Ireland.}

\author{Sam R.~Dolan\,\orcidlink{0000-0002-4672-6523}}
\affiliation{Consortium for Fundamental Physics, School of Mathematics and Statistics, University of Sheffield, Hicks Building, Hounsfield Road, Sheffield S3 7RH, United Kingdom.}

\begin{abstract}
We derive a formalism for solving the Lorenz gauge equations for metric perturbations of Kerr spacetime sourced by an arbitrary stress-energy tensor. The metric perturbation is obtained as a sum of differential operators acting on a set of six scalars, with two of spin-weight $\pm2$, two of spin-weight $\pm1$, and two of spin-weight $0$. We derive the sourced Teukolsky equations satisfied by these scalars, with the sources given in terms of differential operators acting on the stress-energy tensor.
The method can be used to obtain both linear and higher-order nonlinear metric perturbations, and it fully determines the metric perturbation up to a time integral, omitting only static contributions which must be handled separately.
\end{abstract}

\maketitle

\section{Introduction}

Black hole perturbation theory has proved to be a highly effective approach to the two-body problem in general relativity. Waveform models based on black hole-perturbation theory are expected to be a key ingredient in the study of extreme mass-ratio inspirals (EMRIs) by the European Space Agency's forthcoming LISA mission. In the context of black hole perturbation theory, in order to extract the maximum science gain from observations of EMRIs by LISA it is necessary to incorporate effects through second order in perturbation theory \cite{Hinderer:2008dm,Isoyama:2012bx,Burko:2013cca,Burke:2023lno}. This has
recently been achieved in the relatively simple case of a binary where both black holes are non-spinning and the inspiral is quasi-circular \cite{Wardell:2021fyy}. However, it is highly unlikely that LISA will observe such simple EMRIs, so it will be important to incorporate black hole spins, orbital precession and eccentricity into models. For black hole perturbation theory, this translates into the need to solve the second-order linearised Einstein equations on a Kerr background spacetime.

Perturbations of Kerr spacetime are significantly more challenging than those of Schwarzschild spacetime. Notably, the reduced symmetry means that the equations for metric perturbations are not known to admit separable solutions. Teukolsky \cite{Teukolsky:1972my} overcame this problem by instead deriving an equation for certain components of the perturbed Weyl tensor instead of the metric. It turns out that those Teukolsky equations are both decoupled (so one can solve a single equation for a scalar instead of solving 10 coupled equations for 10 components of the metric tensor) and that they admit a separable solution, making it a highly efficient approach to perturbations of Kerr spacetime. However, this introduces another problem: the Teukolsky equations yield solutions representing components of the perturbed Weyl tensor, but many applications require the actual metric perturbation. For example, the first-order metric perturbation is an essential ingredient that appears in the source for the second-order perturbation equations, even when working with a second-order Teukolsky formalism \cite{Spiers:2023cip}.

One solution that achieves the best of both worlds is based on \emph{metric reconstruction}, in which one reconstructs a metric perturbation from solutions of the Teukolsky equation. Chrzanowski, Cohen and Kegeles \cite{Chrzanowski:1975wv,Kegeles:1979an} showed that the metric perturbation in Kerr spacetime can be reconstructed by applying a differential operator to scalars that are related to the Weyl scalars by separable ``inversion relations''. Unfortunately, there are several drawbacks to their reconstruction procedure:
(i) the metric perturbation is in a radiation gauge, which necessarily means that it can’t represent a full solution to a sourced equation unless certain components of the stress-energy tensor are zero \cite{Price:2006ke};
(ii) the “inversion” relation between the Hertz potential and the Weyl tensor requires the solution of a fourth-order equation, introducing technical complexity;
(iii) the reconstructed metric perturbation typically has extended string-like gauge singularities \cite{Barack:2001ph,Ori:2002uv,Keidl:2010pm,Pound:2013faa}. These gauge singularities, in particular, make the metric perturbation unsuitable for constructing a source for the second-order perturbation equations.

In this paper, we develop a metric reconstruction prescription that addresses all of those deficiencies. For alternative, complementary approaches (some of which are not in Lorenz gauge) see Refs.~\cite{Green:2019nam,Loutrel:2020wbw,Ripley:2020xby,Toomani:2021jlo,Osburn:2022bby,Franchini:2023xhd,Bourg:2024vre, Hollands:2024iqp}.

The prescription has three key ingredients that make it an efficient approach to perturbations of Kerr:
\begin{itemize}
    \item It is based on solving (decoupled, separable) Teukolsky equations;
    \item The inversion relations are simple time integrals;
    \item It is in Lorenz gauge.
\end{itemize}
In addition to the obvious benefit of the first two of these, the use of Lorenz gauge brings further advantages:
\begin{itemize}
    \item The singularity arising from a point-particle source is isotropic, and there are well-established numerical methods for handling the singularity in a robust way;
    \item The extended singularities that unavoidably appear in, e.g. radiation gauge, do not appear in Lorenz gauge.
    \item All existing calculations at second order in perturbation theory have made use of Lorenz-gauge;
    \item The asymptotic behaviour towards the horizon and infinity is well-understood and free of divergences;
    \item The equations of motion derived within gravitational self-force theory typically rely on Lorenz gauge for their regularization schemes.
\end{itemize}
The results are a natural extension of earlier work on homogeneous solutions \cite{Dolan:2021ijg} and for the specific source of a point mass on a circular, equatorial orbit of the Kerr black hole \cite{Dolan:2023enf}. The approach to obtaining inhomogeneous solutions is very different to that of Ref.~\cite{Dolan:2023enf}, which fundamentally relied on the properties of the solution for a particle by ``glueing'' together homogeneous solutions at the circular-orbit radius. In addition to being applicable to much more general sources, this new approach also avoids the necessity to project from spheroidal onto spherical harmonic modes and, in fact, does not even require a mode ansatz at all.

This paper is organised as follows. In Sec.~\ref{sec:KerrPerturbations} we review some key results from black hole perturbation theory on which this work is based. In Sec.~\ref{sec:Lorenz} we derive our main result: a set of sourced Teukolsky equations for perturbations of Kerr spacetime in Lorenz gauge. We conclude in Sec.~\ref{sec:Conclusion} with a summary and discussion of future work. In Appendix \ref{sec:irreducible-decompositions} we review some further background material, extending that given in Sec.~\ref{sec:KerrPerturbations}. We introduce a number of operators throughout this paper; the most relevant of these are listed in Table \ref{tab:operators}, and all operators are given explicitly as GHP expressions in Appendix \ref{sec:operators}.
\begin{table}[!ht]
    \centering
    \begin{tabular}{c|l|l}
      & Description & Equation \\ \hline
      $\cE$ & Linearised Einstein & $\cE_{\alpha \beta}{}^{\gamma \delta} (h_{\gamma\delta}) = 8 \pi T_{\alpha \beta}$ \\ \hline
      $\cT_0$ & $s=+2$ Linearised Weyl Scalar & $\psi_0 = \cT_0^{\alpha \beta} (h_{\alpha \beta})$  \\ \hline
      $\cT_4$ & $s=-2$ Linearised Weyl Scalar & $\psi_4 = \cT_4^{\alpha \beta} (h_{\alpha \beta})$ \\ \hline
      $\cS_0$ & $s=+2$ Decoupling & \multirow{2}{*}{$\cO_0 \psi_0 = 8 \pi \cS_0^{\alpha \beta} (T_{\alpha \beta})$} \\
      $\cO_0$ & $s=+2$ Teukolsky & \\ \hline
      $\cS_4$ & $s=-2$ Decoupling & \multirow{2}{*}{$\cO_4 \psi_4 = 8 \pi \cS_4^{\alpha \beta} (T_{\alpha \beta})$} \\
      $\cO_4$ & $s=-2$ Teukolsky &  \\ \hline
      $\cN$ & AAB corrector & $\cN_{\alpha \beta}{}^{\gamma \delta} (T_{\gamma\delta}) $ \\ \hline
      $\cA$ & AAB vector & $\cA_{\alpha}{}^{\beta \gamma} (h_{\beta\gamma}) $ \\ \hline
      $\cT_0$ & $s=+1$ Maxwell Scalar & $\phi_0 = \cT_0^{\alpha \beta} (\xi_{\alpha})$  \\ \hline
      $\cT_2$ & $s=-1$ Maxwell Scalar & $\phi_2 = \cT_2^{\alpha \beta} (\xi_{\alpha})$ \\ \hline
      $\cS_0$ & $s=+1$ Decoupling & \multirow{2}{*}{$\cO_0 \phi_0 = \cS_0^{\alpha} (j_{\alpha})$} \\
      $\cO_0$ & $s=+1$ Teukolsky & \\ \hline
      $\cS_2$ & $s=-1$ Decoupling & \multirow{2}{*}{$\cO_2 \phi_2 = \cS_2^{\alpha} (j_{\alpha})$} \\
      $\cO_2$ & $s=-1$ Teukolsky &
    \end{tabular}
    \label{tab:operators}
    \caption{Summary of operators appearing in this paper and the equation in which they appear. For all of the listed cases the spin and boost weights are equal, $s=b$.}
\end{table}

Throughout this work we follow the conventions of Misner, Thorne and Wheeler
\cite{Misner:1973prb}: a ``mostly positive'' metric signature, $(-,+,+,+)$, is
used for the spacetime metric; the connection coefficients are defined by
$\Gamma^{\lambda}_{\mu\nu}=\frac{1}{2}g^{\lambda\sigma}(g_{\sigma\mu,\nu}
+g_{\sigma\nu,\mu}-g_{\mu\nu,\sigma})$; the Riemann tensor is
$R^{\tau}{}_{\!\lambda\mu\nu}=\Gamma^{\tau}_{\lambda\nu,\mu}
-\Gamma^{\tau}_{\lambda\mu,\nu}+\Gamma^{\tau}_{\sigma\mu}\Gamma^{\sigma}_{\lambda\nu} -\Gamma^{\tau}_{\sigma\nu}\Gamma^{\sigma}_{\lambda\mu}$, the Ricci
tensor and scalar are $R_{\mu\nu}=R^{\tau}{}_{\!\mu\tau\nu}$ and
$R=R_{\mu}{}^{\!\mu}$, and the Einstein equations are
$G_{\mu\nu}=R_{\mu\nu}-\frac{1}{2}g_{\mu\nu}R=8\pi
T_{\mu\nu}$. Standard geometrised units are used, with $c=G=1$.
We use Greek letters for spacetime indices, Latin letters for tetrad indices, and capital letters for spinor indices. Symmetrisation
of indices is denoted using round brackets [e.g. $T_{(\alpha \beta)} = \tfrac12 (T_{\alpha \beta}+T_{\beta
\alpha})$] and anti-symmetrisation using square brackets [e.g. $T_{[\alpha \beta]} = \tfrac12
(T_{\alpha \beta}-T_{\beta \alpha})$], and indices are excluded from symmetrisation by surrounding them by
vertical bars [e.g. $T_{(\alpha | \beta | \gamma)} = \tfrac12 (T_{\alpha \beta \gamma}+T_{\gamma \beta
\alpha})$]. A tensor without indices denotes the trace, e.g. $T = T^\alpha{}_\alpha$, and a caret denotes trace reversal, $\trev{T}_{\alpha \beta} = T_{\alpha\beta} - \frac12 g_{\alpha \beta} T$. The adjoint of an operator is denoted using $\dag$.

\section{Perturbations of Kerr spacetime}
\label{sec:KerrPerturbations}

The sourced Teukolsky equations derived in Sec.~\ref{sec:Lorenz} build on key existing results from black hole perturbation theory. We begin with a review of the most pertinent of those results, presenting them in an operator notation consistent with the rest of this paper. A more detailed review is given in Appendix \ref{sec:irreducible-decompositions} and the references therein.

\subsection{Linearised Einstein equations}

We start by considering an expansion of the metric tensor in a small parameter, $\epsilon$, as
\begin{equation}
  g^{\rm exact}_{\mu\nu} = g_{\mu\nu} + \epsilon h^{(1)}_{\mu\nu} + \epsilon^2 h^{(2)}_{\mu\nu} + O(\epsilon^3).\label{g expansion}
\end{equation}
The metric perturbations $h^{(1)}_{\mu\nu}$, $h^{(2)}_{\mu\nu}$, \ldots, 
all satisfy {\it linear} systems of partial differential equations that take the form
\begin{equation}
(\cE h^{(i)})_{\mu\nu} = S_{\mu\nu}^{(i)}(h^{(i-1)}, \cdots, h^{(1)}, T_{\alpha\beta}),  \label{eq:LEE}
\end{equation}
where
\begin{equation}
    (\cE h)_{\mu\nu} \equiv -\tfrac12\Big[\Box \trev{h}_{\mu \nu} + 2 R^\alpha{}_\mu{}^\beta{}_\nu \trev{h}_{\alpha \beta} + g_{\mu \nu} \nabla_\sigma Z^\sigma - 
2 \nabla_{(\mu} Z_{\nu)}\Big]
\end{equation}
is the linearised Einstein operator
with $Z_\mu \equiv \nabla^\nu \trev{h}_{\mu \nu}$, where $\trev{h}_{\mu \nu} \equiv h_{\mu \nu} - \tfrac{1}{2} g_{\mu \nu} h$ is the trace-reversed metric perturbation. The source $S^{(i)}_{\mu \nu}$ on the right hand side depends on the the stress-energy tensor $T_{\mu\nu}$ and on all lower-order metric perturbations, $h^{(i-1)}, \cdots, h^{(1)}$.
Here, we focus on linear perturbations (for equivalent equations in the non-linear case see, e.g., Refs.~\cite{Spiers:2023cip,Spiers:2023mor}), in which case we drop the $(i)$ superscripts for simplicity and assume $T_{\mu \nu} \sim \epsilon$. Then, the linearised Einstein equation is
\begin{equation}
  (\cE h)_{\mu\nu} = 8 \pi\, T_{\mu\nu}.
\end{equation}
Further specialising to Lorenz gauge, $Z^\mu = 0$, this simplifies to the Lichnerowicz tensor wave equation,
\begin{equation}
    \Box \trev{h}_{\mu \nu} + 2 R^\alpha{}_\mu{}^\beta{}_\nu \trev{h}_{\alpha \beta} = - 16 \pi T_{\mu \nu}. \label{eq:Lichnerowicz}
\end{equation}

\subsection{Teukolsky equations}
Four of the five Weyl scalars ($\psi_i$, $i \in \{0,1,3,4\}$) are zero on the Kerr-NUT background, and hence their linear perturbations are invariant under infinitesimal gauge transformations. Furthermore, $\psi_0$ and $\psi_4$
are invariant under infinitesimal tetrad transformations.
These \emph{maximum spin} scalars are obtained from the metric perturbation by applying the operators $\cT_0$ and $\cT_4$:
\begin{subequations}
\begin{align}
  \psi_0 &= \delta C_{\alpha \beta \gamma \delta} l^\alpha m^\beta l^\gamma m^\delta \equiv \cT_0 h,\\
  \psi_4 &= \delta C_{\alpha \beta \gamma \delta} n^\alpha \mb^\beta n^\gamma \mb^\delta \equiv \cT_4 h.
\end{align}
\end{subequations}
where $\delta C_{\alpha \beta \gamma \delta}$
is the linearised Weyl tensor.

The maximum-spin Weyl scalars satisfy the Teukolsky equations,
\begin{subequations}
\label{eq:teuks2}
\begin{align}
  \cO_0 \psi_0 &= 8 \pi \,\cS_0 T,\\
  \cO_4 \psi_4 &= 8 \pi \,\cS_4 T,
\end{align}
\end{subequations}
where $\cO_0$ and $\cO_4$ are the Teukolsky operators and $\cS_0$ and $\cS_4$ are decoupling operators that give the source to the Teukolsky equation in terms of the stress-energy $T_{\mu\nu}$. (Recall that these operators, along with others, are given as explicit GHP expressions in Appendix \ref{sec:operators}.)

These operators satisfy Wald's operator identities \cite{Wald:1978vm}
\begin{equation}
  \cS_0 \cE h = \cO_0 \cT_0 h, \quad
  \cS_4 \cE h = \cO_4 \cT_4 h,
\end{equation}
along with the adjoint identities
\begin{equation}
  \cE \cS^\dag_0 \zeta^4 \Psi_4 = \cT^\dag_0 \zeta^4 \cO_4 \Psi_4, \quad
  \cE \cS^\dag_4 \zeta^4 \Psi_0 = \cT^\dag_4 \zeta^4 \cO_0 \Psi_0,
\end{equation}
where the first identity is understood to act on objects $\Psi_4$ of GHP type $\{-4,0\}$ (the same as $\psi_4$), and where the second identity is understood to act on objects $\Psi_0$ of GHP type $\{4,0\}$ (the same as $\psi_0$). They also satisfy the operator identities \cite{Pound:2021qin,Hollands:2024iqp}
\begin{subequations}
\label{eq:TSdag}
\begin{align}
 \cT_0 \cS_0^\dag \zeta^4 \Psi_4 &= 0, \\
 \cT_0 \overline{\cS_0^\dag \zeta^4 \Psi_4}&= \frac14 \th^4 \bar{\zeta}^4 \bar{\Psi}_4, \\
 \cT_4 \cS_0^\dag \zeta^4 \Psi_4&= -\frac34 M \LieT \Psi_4 \nonumber \\
 & \hspace{-3em}+ \frac14[\zeta^4 \cO_4 - 4 (\rho \th' - \tau \edth') \zeta^4 + 8 \psi_2 \zeta^4] \cO_4 \Psi_4, \\
 \cT_4 \overline{\cS_0^\dag \zeta^4 \Psi_4} &= \frac14 \edth'^4 \bar{\zeta}^4 \bar{\Psi}_4, \\
 \cT_0 \cS_4^\dag \zeta^4 \Psi_0 &= \frac34 M \LieT \Psi_0 \nonumber \\
 & \hspace{-3em}+ \frac14[\zeta^4 \cO_0 - 4 (\rho' \th - \tau' \edth) \zeta^4 + 8 \psi_2 \zeta^4] \cO_0 \Psi_0, \\
 \cT_0 \overline{\cS_4^\dag \zeta^4 \Psi_0}&= \frac14 \edth^4 \bar{\zeta}^4 \bar{\Psi}_0, \\
 \cT_4 \cS_4^\dag \zeta^4 \Psi_0&= 0,\\
 \cT_4 \overline{\cS_4^\dag \zeta^4 \Psi_0} &= \frac14 \th'^4 \bar{\zeta}^4 \bar{\Psi}_0,
\end{align}
\end{subequations}
along with the adjoint identities
\begin{subequations}
\label{eq:STdag}
\begin{align}
 \cS_0 \cT_0^\dag \Psi_4 &= 0, \\
 \cS_0 \overline{\cT_0^\dag \Psi_4}&= \frac14 (\th-\rho-\bar{\rho})^4 \bar{\Psi}_4, \\
 \cS_0 \cT_4^\dag \Psi_0&= \frac34 M \zeta^{-4} \LieT \Psi_0 \nonumber \\
 & \hspace{-3em}+ \frac14 \cO_0\big[\cO_0 + 4 \big(\rho (\th'-\rho') - \tau (\edth'-\tau')\big) + 4 \psi_2\big] \Psi_0, \\
 \cS_0 \overline{\cT_4^\dag \Psi_0} &= \frac14 (\edth-\tau-\bar{\tau}')^4 \bar{\Psi}_0, \\
 \cS_4 \cT_0^\dag \Psi_4 &= -\frac34 M \zeta^{-4} \LieT \Psi_4 \nonumber \\
 & \hspace{-3em}+ \frac14\cO_4\big[\cO_4  + 4 \big(\rho' (\th-\rho) - \tau' (\edth-\tau)\big) +4 \psi_2 \big] \Psi_4, \\
 \cS_4 \overline{\cT_0^\dag \Psi_4}&= \frac14 (\edth'-\tau'-\bar{\tau})^4 \bar{\Psi}_4, \\
 \cS_4 \cT_4^\dag \Psi_0&= 0,\\
 \cS_4 \overline{\cT_4^\dag \Psi_0} &= \frac14 (\th'-\rho'-\bar{\rho}')^4  \bar{\Psi}_0.
\end{align}
\end{subequations}

%which we do not require here but are given in, e.g. \cite[Eq.~(K.2)]{Hollands:2024iqp}.

\subsection{Aksteiner, Andersson and B\"ackdahl metric perturbation}

Aksteiner, Andersson and B\"ackdahl (AAB)  \cite{Aksteiner:2016pjt} derived the operator identity (see also \cite[Eq. (K.6)]{Hollands:2024iqp})
\begin{align}\label{eq:AAB-op-id}
  &(M \LieT h)_{\alpha \beta} = \nonumber \\
  & \quad  \Big[\Big(\frac43\cS_4^\dag \zeta^4 \cT_0 - \frac43\cS_0^\dag \zeta^4 \cT_4 + \cN \cE\Big) h\Big]_{\alpha \beta} -2 \nabla_{(\alpha} (\cA h)_{\beta)},
\end{align}
where
\begin{equation}
\cN T = \frac{1}{3}\Big[\cC^\dag \zeta^4 \cK^1 (\cP^{3/2}-\cP^{1/2}) \cC - 3 \psi_2 \zeta^4 \cK^1\Big]S
\end{equation}
with $S_{\mu \nu} = T_{\mu \nu} - \frac14 g_{\mu\nu} T$ a symmetric, trace-free tensor, and where $\cA$ is a third-order differential operator. The curl ($\cC$), curl-dagger ($\cC\dagger$), spin-projection ($\cP^i$) and sign-flipping ($\cK^1$) operators are defined in Appendix \ref{sec:irreducible-decompositions}, as is the Killing spinor coefficient, $\zeta$. There is also the corresponding adjoint identity
\begin{align}\label{eq:AAB-op-id-adj}
  &(M \LieT T)_{\alpha \beta} = \nonumber \\
  & \,\,\, - \Big[\Big(\frac43 \cT_0^\dag \zeta^4 \cS_4 - \frac43 \cT_4^\dag \zeta^4 \cS_0 - \cE \cN + 2 \cA^\dag \Div\Big)
  T\Big]_{\alpha \beta}
\end{align}
where $\cA^\dag$ is the adjoint of $\cA$.

Both operator identities hold when acting on an arbitrary symmetric rank-2 tensor.
Note that since $\cS_4^\dag$ and $\cT_4^\dag$ only have $ll$, $lm$ and $mm$ components, and $\cS_0^\dag$ and $\cT_0^\dag$ only have $nn$, $n\mb$ and $\mb\mb$ components, the tetrad components of these identities decouple:
\begin{subequations}
\label{eq:AAB-op-id-adj-comp}
\begin{align}
  &\{nn, n \mb, \mb\mb\}:\\
  &\qquad\Big[\Big(\cE \cN + \frac43 \cT^\dag_4 \zeta^4 \cS_0 - 2 \cA^\dag \Div\Big)T\Big]_{\alpha \beta} = (M \LieT T)_{\alpha \beta},\nonumber \\
  &\{ll, lm, mm\}:\\
  &\qquad\Big[\Big(\cE \cN -\frac43 \cT^\dag_0 \zeta^4 \cS_4 - 2 \cA^\dag \Div\Big)T\Big]_{\alpha\beta} = (M \LieT T)_{\alpha \beta}, \nonumber \\
  &\{ln, m\mb, l \mb, n m\}: \\
  &\qquad\Big[\Big(\cE \cN-2 \cA^\dag \Div\Big)T\Big]_{\alpha \beta} = (M \LieT T)_{\alpha \beta}.\nonumber
\end{align}
\end{subequations}
These are understood as tensor identities that apply for the components listed in the curly braces.

Finally, related to the identities \eqref{eq:AAB-op-id} and \eqref{eq:AAB-op-id-adj} there are the operator identities
\begin{subequations}
\label{eq:TN}
\begin{align}
 \cT_0 \cN T &= -\frac13[\zeta^4 \cO_0 - 4 (\rho' \th - \tau' \edth) \zeta^4 + 8 \psi_2 \zeta^4] \cS_0 T \nonumber \\
 & \quad
+ \cB_0 \Div T , \\
 \cT_4 \cN T &= \frac13[\zeta^4 \cO_4 - 4 (\rho \th' - \tau \edth') \zeta^4 + 8 \psi_2 \zeta^4] \cS_4 T \nonumber \\
 & \quad
 + \cB_4 \Div T.
\end{align}
\end{subequations}
along with the adjoint identities
\begin{subequations}
\label{eq:NT}
\begin{align}
 \cN &\cT_0^\dag \Psi_4 = - \nabla_{(\alpha} (\cB_0^\dag \Psi_4)_{\beta)}  \nonumber \\
  & +\frac13 \cS_0^\dag\zeta^4\big[ \cO_4 + 4 \big(\rho' (\th-\rho) - \tau' (\edth-\tau)\big) + 4 \psi_2\big] \Psi_4, \\
 \cN &\cT_4^\dag  \Psi_0 =  - \nabla_{(\alpha} (\cB_4^\dag \Psi_0)_{\beta)} \nonumber \\
 & -\frac13\cS_4^\dag\zeta^4 \big[\cO_0 + 4 \big(\rho (\th'-\rho') - \tau (\edth'-\tau')\big) + 4 \psi_2\big]  \Psi_0.
\end{align}
\end{subequations}
Here, $\cB_0$ and $\cB_4$ are second-order differential operators and $\cB_0^\dag$ and $\cB_4^\dag$ are their adjoints.

We now define a metric perturbation
\begin{align}\label{eq:AAB}
  h^\AAB_{\alpha \beta} 
  & = -\frac23 \nabla^\mu \zeta^4 \nabla^\nu \cC^{-}_{\mu(\alpha|\nu|\beta)} +8\pi (\cN T)_{\alpha \beta}
  \nonumber \\
   &\quad= \frac43 (\cS_4^\dag \zeta^4 \psi_0 - \cS_0^\dag \zeta^4 \psi_4)_{\alpha \beta} +8\pi (\cN T)_{\alpha \beta},
\end{align}
which is just Eq.~\eqref{eq:AAB-op-id} with the Lie derivative absorbed into the definition on the left-hand side, with the replacements $\cT_0 h \to \psi_0$, $\cT_4 h \to \psi_4$ and $(\cE h)_{\alpha \beta} \to 8 \pi T_{\alpha \beta}$ on the right-hand side, and with the omission of the gauge term involving the vector $(\cA h)_\alpha$.
This metric perturbation is trace-free ($h^\AAB = 0$) and it satisfies
\begin{equation} \label{eq:ddhAAB}
  \nabla^\beta \nabla^\alpha h^\AAB_{\alpha \beta} = 8\pi\nabla^\beta \nabla^\alpha (\cN T)_{\alpha \beta} = - 8 \pi M \LieT T.
\end{equation}
Provided $\psi_0$ and $\psi_4$ are the perturbed Weyl scalars corresponding to a solution to the linearised Einstein equation (or, equivalently, that they satisfy the Teukolsky equations \eqref{eq:teuks2}) then this ``AAB'' metric perturbation is a complex solution of the linearised Einstein equation with a source which is the time derivative of the stress energy:
\begin{equation}
  (\cE h^\AAB)_{\mu\nu} = 8 \pi M \LieT T_{\mu\nu}.
\end{equation}
This is a consequence of the operator identity \eqref{eq:AAB-op-id} (or, equivalently, the operator identity \eqref{eq:AAB-op-id-adj}).
Since $\cE$, $\LieT$ and $T_{\mu \nu}$ are all real, the complex conjugate of the AAB metric perturbation is also a solution of the same equation, and we can take the real part to obtain a real metric perturbation.

As expected from the fact that the AAB metric perturbation is the time derivative of a solution of the Einstein equation, the identities \eqref{eq:TSdag} and \eqref{eq:TN} along with the Teukolsky equations for $\psi_0$ and $\psi_4$ imply that the Weyl scalars derived from it satisfy ``circularity relations''
\begin{subequations}
\begin{align}
  \cT_0 h^\AAB &= M \LieT \psi_0,\\
  \cT_4 h^\AAB &= M \LieT \psi_4.
\end{align}
\end{subequations}
If we instead consider the complex conjugate of the AAB metric perturbation then we find
\begin{subequations}
\begin{align}
  \cT_0 \overline{h^\AAB} &= \frac13 (\edth^4 \bar{\zeta}^4 \bar{\psi}_0 - \th^4 \bar{\zeta}^4 \bar{\psi}_4) + 8 \pi \cT_0 \bar{\cN} T,\\
  \cT_4 \overline{h^\AAB} &= -\frac13 (\edth'^4 \bar{\zeta}^4 \bar{\psi}_4-\th'^4 \bar{\zeta}^4 \bar{\psi}_0)  + 8 \pi \cT_4 \bar{\cN} T.
\end{align}
\end{subequations}
Combining these, we obtain Teukolsky-Starobinsky identities for sourced perturbations \cite{Aksteiner:2016mol,Hollands:2024iqp}:
\begin{subequations}
\begin{align}
  \th^4 \zeta^4 \psi_4  &= \edth'^4 \zeta^4 \psi_0 - 3 M \LieT \bar{\psi}_0 + 24 \pi \bar{\cT}_0 \cN T, \\
  \th'^4 \zeta^4 \psi_0 &= \edth^4 \zeta^4 \psi_4 + 3 M \LieT \bar{\psi}_4 - 24 \pi \bar{\cT}_4 \cN T.
\end{align}
\end{subequations}

\section{Lorenz gauge metric perturbation}
\label{sec:Lorenz}

The AAB solution is not in Lorenz gauge. We now transform to Lorenz gauge by making a gauge transformation,
\begin{equation} \label{eq:h-Lor}
 M \LieT h^\Lor_{\alpha \beta} = h^\AAB_{\alpha\beta} - 2 \xi_{(\alpha; \beta)}.
\end{equation}
such that
\begin{equation}\label{eq:lorenz}
  M \LieT \nabla^\alpha \trev{h}^L_{\alpha \beta} = 0.
\end{equation}
This implies that the gauge vector $\xi_\alpha$ must satisfy
\begin{equation}
\label{eq:vector-wave}
    \Box \xi_\alpha = \nabla^\beta h^{\rm AAB}_{\alpha \beta}.
\end{equation}
We write the solution to this equation as a sum of pieces
and will derive equations for each of the pieces.

\subsection{Trace piece}
We first seek to ensure that the trace of the metric perturbation is in Lorenz gauge. This trace piece satisfies the trace of Eq.~\eqref{eq:Lichnerowicz},
\begin{align}
\label{eq:hL}
    \Box h^\Lor = 16 \pi T.
\end{align}
Introducing the gauge vector
\begin{equation}
    \xi_\alpha^\trace = \frac12 f_\alpha{}^\beta h^\Lor_{;\beta} + \kappa_{;\alpha},
\end{equation}
and recalling that the AAB metric perturbation is trace-free, $h^\AAB = 0$, we obtain an equation for $\kappa$ by requiring $-2\nabla^\alpha \xi_\alpha^\trace = M \LieT h^\Lor$,
\begin{equation}
\label{eq:kappa}
    \Box \kappa = M \LieT h^\Lor.
\end{equation}
A straightforward calculation shows that we then have
\begin{subequations}
\begin{align}
    \Box \xi^\trace_{\alpha} &= 8 \pi \big[\nabla_\beta(f_\alpha{}^\beta T) - \TKV_\alpha T \big] \equiv j_\alpha^\trace, \label{eq:vector-wave-tr}\\
    \nabla^\alpha j_\alpha^\trace &= - 8 \pi M\LieT T \label{eq:div-j-tr}.
\end{align}
\end{subequations}

\subsection{Trace-free piece}
By construction, the remainder of the gauge vector,
\begin{equation}
    \xi_\alpha^\maxwell = \xi_\alpha - \xi_\alpha^\trace,
\end{equation}
satisfies $\nabla^\alpha \xi_\alpha^\maxwell = 0$. Equations \eqref{eq:vector-wave} and \eqref{eq:vector-wave-tr} imply that $\xi_\alpha^\maxwell$ satisfies the vector wave equation,
\begin{equation}
    \Box \xi_\alpha^\maxwell = j_\alpha = j_\alpha^\Weyl + j_\alpha^\SE,
\end{equation}
with source given by
\begin{subequations}
\begin{align}
     j_\alpha^\Weyl &= -\frac23 \nabla^\beta \nabla^\mu \zeta^4 \nabla^\nu \cC^{-}_{\mu(\alpha|\nu|\beta)} = \nabla^\beta \cJ^\Weyl_{\alpha \beta}, \\
     j_\alpha^\SE &= \nabla^\beta (\cN T)_{\alpha \beta} - j_\alpha^\trace,
\end{align}
\end{subequations}
where
\begin{equation}
    \cJ^\Weyl_{\alpha \beta} = \frac43 \nabla^\gamma \zeta^{-1} \nabla^\delta \zeta^5 \cC^-_{\gamma \delta \alpha \beta} = - \frac43 \nabla^\gamma (U^\delta \zeta^4 \cC^-_{\gamma \delta \alpha \beta})
\end{equation}
and $U_\gamma = \nabla_\gamma(\log \zeta)$.
Note that $\cJ^\Weyl_{\alpha \beta}$ has components
\begin{subequations}
\label{eq:JWeyl}
\begin{align}
    \cJ^\Weyl_{\mb n} &= -\frac43 (\rho \edth - \tau \th) \zeta^{4} \psi_{4},\\
    \cJ^\Weyl_{lm} &= -\frac43 (\rho' \edth' - \tau' \th') \zeta^{4} \psi_{0}.
\end{align}
\end{subequations}

Furthermore, Eqs.~\eqref{eq:ddhAAB} and \eqref{eq:div-j-tr} imply that the two pieces of the source are separately conserved, $\nabla^\alpha j_\alpha^\Weyl = 0 = \nabla^\alpha j_\alpha^\SE$, so we have a Maxwell-type problem in vector Lorenz gauge for $\xi_\alpha^\maxwell$. We can solve this using a circularity reformulation of the methods of Green and Toomani \cite{Green-talk-2021,Green-Toomani} and Dolan, Kavanagh and Wardell (DKW) \cite{Dolan:2021ijg,Dolan:2023enf}. In doing so, it is convenient to further split $\xi^\maxwell_\alpha$ into a piece sourced by $\cC_{\alpha \beta \gamma \delta}$ and a piece sourced only by the stress-energy,
\begin{equation}
    \xi^\maxwell_\alpha = \xi^\Weyl_\alpha + \xi^\SE_\alpha.
\end{equation}

\subsubsection{Weyl-sourced piece}

We can use the existing DKW gauge vector for source free perturbations \cite{Dolan:2021ijg} to solve for $\xi^\Weyl_\alpha$ by writing it as
\begin{equation}
\xi^\Weyl_\alpha = \xi^\DKW_\alpha - \nabla_\alpha \chi^{\DKWSE}
\end{equation}
where
\begin{equation}
\xi^\DKW_\alpha = \zeta^2 \nabla^\beta \cH_{\alpha \beta}^\DKW - \nabla_\alpha \chi^\DKW
\end{equation}
with
\begin{equation}
  \label{eq:H-DKW}
    \LieT \cH_{\alpha \beta}^\DKW =  \frac{4}{9\zeta^2}(l_{[\alpha} m_{\beta]}\cH_4 \zeta^4 \psi_4 - \mb_{[\alpha} n_{\beta]} \cH_0 \zeta^4 \psi_0 ),
\end{equation}
and
\begin{equation}
  \label{eq:chi-DKW}
    \LieT^2 \chi^\DKW = \frac{1}{18}(\chi_4  \zeta^4 \psi_4 - \chi_0  \zeta^4 \psi_0).
\end{equation}
Here, we have introduced the operators $\chi_0$, $\chi_4$, $\cH_0$, $\cH_4$, which are defined by
\begin{subequations}
\begin{alignat}{3}
\chi_0 & = \edth'^2 \bar{\zeta^2} \th'^2, &\quad 
\cH_0  & = \edth' \bar{\zeta}\th',\\
\chi_4 & = \edth^2 \bar{\zeta^2} \th^2, &
\cH_4  & = \edth \bar{\zeta}\th.
\end{alignat}
\label{eq:H-operators}
\end{subequations}
In the case where $\psi_0$ and $\psi_4$ satisfy \textit{sourced} Teukolsky equations, the DKW gauge vector on its own does not satisfy the homogeneous Lorenz gauge equation or the Lorenz gauge condition,
\begin{align}
      \Box \xi_\alpha^\DKW - j_\alpha^\Weyl \ne 0 , \qquad
      \nabla^\alpha \xi_\alpha^{\DKW} \ne 0,
\end{align}
nor does it satisfy the homogeneous Maxwell equation,
\begin{align}
      2 \nabla^\beta \nabla_{[\beta} \xi^\DKW_{\alpha]} - j^\Weyl_\alpha \equiv j_\alpha^{\DKWSE} &\ne 0.
\end{align}
The source $j_\alpha^{\DKWSE}$ has tetrad components
\begin{subequations}
\begin{align}
  M \LieT j_\alpha^{\DKWSE} &= \frac{4}{9\zeta} (n_\alpha \edth' - \mb_\alpha \th') \zeta \cH_0^\dag \zeta^4 (\cS_0 T) \nonumber \\
  & \quad +\frac{4}{9\zeta} (l_\alpha \edth - m_\alpha \th) \zeta \cH_4^\dag  \zeta^4 (\cS_4 T),
\end{align}
\end{subequations}
and the Lorenz gauge violation is given by
\begin{align}
  (M \LieT)^2 \nabla^\alpha \xi_\alpha^{\DKW} =
   \,\frac{1}{9} \Big[\chi_4^\dag \zeta^4 (\cS_4 T) - \chi_0^\dag \zeta^4 (\cS_0 T)\Big].
\end{align}
Note that we have written these such that they manifestly involve the adjoints of the operators appearing in Eqs.~\eqref{eq:H-DKW} and \eqref{eq:chi-DKW}. We also note the operator identities\footnote{The last two of these follow as a trivial consequence of the fact that the spin-1 Teukolsky operator can be written in factorised form \cite{Wardell:2020naz} (see also Eqs.~(3.39) and (3.43) of Ref.~\cite{Aksteiner:2014zyp}):
\begin{subequations}
\begin{align*}
    \cO_0 \phi_{0} &= 4\cS_{0} \zeta^{-2} \cS^\dag_{2} \zeta^{2} \phi_{0}, \\
    \cO_2 \phi_{2} &= 4\cS_{2} \zeta^{-2}\cS^\dag_{0}\zeta^{2} \phi_{2}.
\end{align*}
\end{subequations}
The spin-1 decoupling operator $\cS_0$ and the Teukolsky operator $\cO_0$ here are not to be confused with the spin-2 decoupling operator and the Teukolsky operator appearing in Eq.~\eqref{eq:teuks2}. There is no ambiguity as the choice of whether they represent the spin-1 or spin-2 operators is clear from the object on which they are acting.}
\begin{equation}
  \th^n\zeta^{n} \edth'^n = \edth'^n \zeta^{n} \th^n,
\end{equation}
\begin{align}
   (\cO \chi_4 \zeta^4 - \chi_4^\dagger \zeta^4 \cO_4) \psi_4 &- (\cO \chi_0 \zeta^4 - \chi_0^\dagger \zeta^4 \cO_0) \psi_0 \nonumber \\
   &= 3 M \LieT \nabla_\beta \big(f^{\beta \alpha} j^\Weyl_\alpha) ,
\end{align}
\begin{align}
    2\cS_{0}^\beta (\nabla^\alpha \cJ_{\alpha\beta}) &= \cO_0 \cJ_{l m}, \\
    2\cS_{2}^\beta (\nabla^\alpha \cJ_{\alpha\beta}) &= \cO_2 \cJ_{\mb n},
\end{align}
which are useful in deriving these expressions.

By demanding that $\nabla^\alpha \xi^\Weyl_\alpha = 0$ we then get an equation for $\chi^{\DKWSE}$:
\begin{align}
\label{eq:chi-DKWT}
    \Box \chi^{\DKWSE} &= \nabla^\alpha \xi_\alpha^{\DKW}.
\end{align}
Since this is an equation sourced only by the stress-energy, it is natural to incorporate it into the stress-energy sourced piece, and we will ultimately do so in the final grouping of terms given in Sec.~\ref{sec:spin-decomp}.

\subsubsection{Stress-energy sourced piece: solution via circularity}

The remaining piece of the gauge vector satisfies a compactly supported vector wave equation in Lorenz gauge,
\begin{equation}
    \Box \xi_\alpha^\SE = j_\alpha^\SE - j_\alpha^\DKWSE.
\end{equation}
This can be efficiently solved using a circularity formulation in which the solution is given by
\begin{equation}
    \xi^\SE_\alpha = \nabla^{\beta}\cH^\SE_{\alpha\beta} + f_\alpha{}^\beta \cJ^\SE_\beta - \nabla_\alpha \chi^\SE,
\end{equation}
where
\begin{equation}
    M\LieT \cJ^\SE_\beta = j_\beta^\SE - j_\alpha^\DKWSE
\end{equation}
is chosen such that $\cH^\SE_{\alpha\beta}$ satisfies the circularity relation 
\begin{equation}
    M \LieT \cH^\SE_{\alpha \beta} = f_{[\alpha}{}^\gamma \cF^\SE_{|\gamma| \beta]}, \quad \cF^\SE_{\alpha \beta} = 2 \xi^\SE_{[\beta;\alpha]},
\end{equation}
and where we require
\begin{equation}
\label{eq:teuk-s0}
    \Box \chi^\SE = \nabla_\alpha \big(f^{\alpha \beta} \cJ_\beta\big)
\end{equation}
to enforce the Lorenz gauge condition, $\nabla^\alpha \xi^\SE_\alpha = 0$. The tetrad components of the two-form $\cH^\SE_{\alpha \beta}$ satisfy the spin $\pm1$ Teukolsky equations,
\begin{subequations}
\begin{align}
    \cO_0 \phi_0^\SE &= \cS_0 (j^\SE - j^\DKWSE) \nonumber \\
    & =  \cS_0 j^\SE - \frac{2}{9} (M \LieT)^{-2} \cO \cH^\dag_0 \zeta^4  (\cS_0 T)\\
    \cO_2 \phi_2^\SE &= \cS_2 (j^\SE - j^\DKWSE) \nonumber \\
    & =  \cS_2 j^\SE + \frac{2}{9} (M \LieT)^{-2} \cO_2 \cH^\dag_4 \zeta^4 (\cS_4 T).
\end{align}%
\label{eq:teuk-s1}%
\end{subequations}
where $\cH^\dag_0$ and $\cH^\dag_4$ are adjoints of the operators defined in Eq.~\eqref{eq:H-operators}.

\subsection{Spin decomposition}
\label{sec:spin-decomp}

In summary, we have obtained a Lorenz gauge metric perturbation that satisfies (up to a time derivative) the linearised Einstein equation with a source. The metric perturbation is given by Eq.~\eqref{eq:h-Lor} along with Eq.~\eqref{eq:AAB} and the gauge vector $\xi^\alpha$ that transforms to Lorenz gauge. Gathering the final result together, we have
\begin{equation}
 M \LieT h^\Lor_{\alpha \beta} = -\frac23 \nabla^\mu \zeta^4 \nabla^\nu \cC^{-}_{\mu(\alpha|\nu|\beta)} + (\cN T)_{\alpha \beta} - 2 \xi_{(\alpha; \beta)}
\end{equation}
with
\begin{gather}
    \xi_\alpha = \zeta^2 \nabla^\beta \cH_{\alpha \beta}^\DKW + \nabla^{\beta}\cH^\SE_{\alpha\beta}
    + f_\alpha{}^\beta (\cJ^\SE_\beta + \frac12 \nabla_\beta h^\Lor)\nonumber \\
    - \nabla_\alpha(\chi^\DKW + \chi^{\DKWSE} + \chi^\SE - \kappa).
\end{gather}

The metric perturbation is given in terms of differential operators acting on a set of six Teukolsky scalars, two of spin-2 ($\psi_0$ and $\psi_4$), two of spin-1 ($\phi_0^\SE$ and $\phi_2^\SE$), and two of spin-0 ($h^\Lor$ and the combination $\chi := \chi^\DKWSE + \chi^\SE - \kappa$). These satisfy six sourced Teukolsky equations: Eqs.~\eqref{eq:teuks2} for $\psi_0$ and $\psi_4$; Eqs.~\eqref{eq:teuk-s1} for $\phi_0^\SE$ and $\phi_2^\SE$; Eq.~\eqref{eq:hL} for $h^\Lor$; and Eqs.~\eqref{eq:kappa}, \eqref{eq:chi-DKWT} and \eqref{eq:teuk-s0} for the scalars appearing in $\chi$. Collecting those equations together, we must solve the six Teukolsky equations
\begin{subequations}
\begin{align}
  \cO_0 \psi_0 &= 8 \pi \,\cS_0 T,\\
  \cO_4 \psi_4 &= 8 \pi \,\cS_4 T,\\
  \cO_0 \phi_0^\SE &= \cS_0 j^\SE - \frac{2}{9} (M \LieT)^{-2} \cO_0 \cH^\dag_0 \zeta^4  (\cS_0 T)\\
  \cO_2 \phi_2^\SE &= \cS_2 j^\SE + \frac{2}{9} (M \LieT)^{-2} \cO_2 \cH^\dag_4 \zeta^4 (\cS_4 T)\\
  \Box h^\Lor &= 16 \pi T \\
  \Box \chi &=  (M\LieT)^{-1} \nabla_\alpha \big(f^{\alpha \beta} j^\SE_\beta\big) - M \LieT h^\Lor \nonumber \\
  & \quad - (M \LieT)^{-2} \frac{1}{3} \Big[\chi_4 \zeta^4 (\cS_4 T)
  - \chi_0 \zeta^4 (\cS_0 T)\Big].
\end{align}
\end{subequations}

It is convenient to rearrange the metric reconstruction expression, grouping terms by the spin-weight of scalar from which they are derived:
\begin{equation}
    \LieT h^\Lor_{\alpha \beta} = h^{(s=2)}_{\alpha \beta} + h^{(s=1)}_{\alpha \beta} + h^{(s=0)}_{\alpha \beta} + h^{\SE}_{\alpha \beta},
\end{equation}
where
\begin{subequations}
    \begin{align}
        h^{(s=2)}_{\alpha \beta} &= -\frac23 \nabla^\mu \zeta^4 \nabla^\nu \cC^{-}_{\mu(\alpha|\nu|\beta)} \nonumber \\
        & \quad  -2 \nabla_{(\alpha} \big[\zeta^2 \nabla^\gamma \cH_{\beta) \gamma}^\DKW\big] + 2 \nabla_\alpha \nabla_\beta \chi^\DKW, \\
        h^{(s=1)}_{\alpha \beta} &= -2\nabla_{(\alpha}\nabla^{\gamma}\cH^\SE_{\beta)\gamma}, \\
        h^{(s=0)}_{\alpha \beta} &= - \nabla_{(\alpha}f_{\beta)}{}^\gamma \nabla_\gamma h^\Lor \nonumber \\
        & \quad  + 2 \nabla_\alpha \nabla_\beta (\chi^{\DKWSE} + \chi^\SE - \kappa), \\
        h^{\SE}_{\alpha \beta} &= (\cN T)_{\alpha \beta} - 2 \nabla_{(\alpha} \big[f_{\beta)}{}^\gamma \cJ^\SE_\gamma \big].
    \end{align}
\end{subequations}

\subsection{Solutions to sourced Teukolsky equations}
We have reduced the problem of solving the linearised Einstein equation for the metric perturbation to the problem of solving six sourced Teukolsky equations. The solutions to these are readily obtained via the Green function method, which reduces to the standard variation-of-parameters method when working with a decomposition into mode solutions of the radial Teukolsky equations. Then, since we have written everything in terms of operators we can write the solutions in terms of the adjoints of those operators, i.e.~for a scalar $\Psi(x)$ and a Green function $G(x,x')$ for the equation satisfied by $\Psi(x)$ we have
\begin{align}
    \Psi(x) %&= \int G^\dag(x,x') \cS T \sqrt{-g}\, d^4 x \nonumber \\
    &=  \int (\cS^\dagger G)^{\alpha \beta} T_{\alpha \beta} \sqrt{-g}\, d^4 x'
\end{align}
Applying this to the case at hand, we get
\begin{widetext}\begin{subequations}
\label{eq:solutions}
\begin{align}
  \psi_0 &= 8 \pi \int T_{\alpha \beta} (\cS_0^\dag G_0)^{\alpha \beta} \sqrt{-g}\, d^4 x',\\
  \psi_4 &= 8 \pi \int T_{\alpha \beta} (\cS_4^\dag G_4)^{\alpha \beta} \sqrt{-g}\, d^4 x',\\
  \phi_0^\SE &= \int T_{\gamma \delta}\big((8 \pi f_{\alpha\beta} g^{\gamma \delta} -\cN^\dag_{\alpha \beta}{}^{\gamma \delta}) \nabla^\beta + 8 \pi g^{\gamma \delta} T_\alpha\big)(\cS_0^\dag G_0)^{\alpha} \sqrt{-g}\, d^4 x' - \frac{2}{9} (M \LieT)^{-2} \cH^\dag_0 \zeta^4  (\cS_0 T),\\
  \phi_2^\SE &= \int T_{\gamma \delta}\big((8 \pi f_{\alpha\beta} g^{\gamma \delta} -\cN^\dag_{\alpha \beta}{}^{\gamma \delta}) \nabla^\beta + 8 \pi g^{\gamma \delta} T_\alpha\big)(\cS_2^\dag G_2)^{\alpha} \sqrt{-g}\, d^4 x' + \frac{2}{9} (M \LieT)^{-2} \cH^\dag_4 \zeta^4 (\cS_4 T),\\
  h^\Lor &= 16 \pi \int G(x,x') T \sqrt{-g}\, d^4 x', \\
  \chi &= \int \big[(M\LieT)^{-1} T_{\gamma \delta} \big((8 \pi f_{\alpha\beta} g^{\gamma \delta} -\cN^\dag_{\alpha \beta}{}^{\gamma \delta}) \nabla^\beta + 8 \pi g^{\gamma \delta} T_\alpha\big)f^{\gamma \alpha} \nabla_\gamma G(x,x') \nonumber \\
  & \qquad - G(x,x') M \LieT h^\Lor - \frac13 (M\LieT)^{-2}T_{\alpha \beta} \big(\cS^{\dag\, \alpha \beta}_{4} \zeta^4 \chi^\dag_4 - \cS^{\dag\, \alpha \beta}_{0} \zeta^4 \chi^\dag_0 \big)G(x,x')\big]\sqrt{-g}\, d^4 x'.
\end{align}
\end{subequations}
\end{widetext}

\begin{figure}[htb!]
    \centering
    \includegraphics[width=\columnwidth]{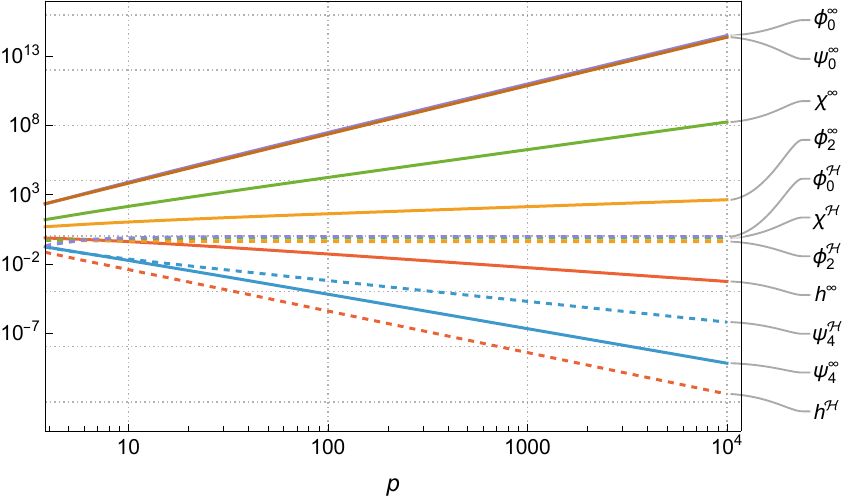}
    \caption{Absolute value of the asymptotic amplitudes of the $(\ell,m)=(2,2)$ spin-weighted spheroidal harmonic mode of the solutions to Eqs.~\eqref{eq:solutions} for a point particle an a circular orbit of radius $r_0=p M$ in Kerr spacetime with $a=0.6M$. The coefficients of the dominant behaviour towards infinity are shown by solid lines and towards the horizon by dashed lines.}
    \label{fig:amplitudes}
\end{figure}
As a check of these expressions, we have implemented them in the Teukolsky \cite{TeukolskyBHPT} package of the Black Hole Perturbation Toolkit. Our code will be released publicly along with a follow-on paper describing the implementation for the case of a point particle on a generic, bound orbit in Kerr spacetime \cite{Cunningham-Lorenz}. In short, for each of the six fields in $\Psi \in (\psi_0,\psi_4, \phi_0^\SE, \phi_2^\SE, h^\Lor,\chi^{\DKWSE}+\chi^\SE)$, our implementation considers their decomposition into spin-weighted spheroidal harmonic modes,
\begin{equation}
   \Psi = \int_{-\infty}^\infty \sum_{\ell=|s|}^\infty \sum_{m=-\ell}^\ell \frac{\Psi^{\ell m \omega}(r)}{\zeta^{|s|-s}} \, {}_s S_{\ell m}(\theta, \phi; a \omega) e^{-i \omega t} d\omega , \label{eq:Psi-FD}
\end{equation}
where the functions $\Psi^{\ell m \omega}(r)$ and ${}_{s} S_{\ell m}(\theta, \phi; a \omega)$ satisfy the radial Teukolsky and spin-weighted spheroidal harmonic equations, respectively. We then use variation of parameters to determine the radial function in terms of a linear combination of unit-normalised ``in'' and ``up'' homogeneous solutions \cite{Pound:2021qin},
\begin{equation} \label{eq:TeukolskyInhomogeneousModes}
\Psi^{\ell m \omega}(r) = 
\begin{cases}
\Psi^{\cH}{}_s R^{\text{in}}_{\ell m \omega}(r) & r \le r_0,\\
\Psi^{\infty} {}_s R^{\text{up}}_{\ell m \omega}(r) & r\ge r_0,
\end{cases}
\end{equation}
where for simplicity we have suppressed the fact that the \emph{asymptotic amplitudes} $\Psi^{\cH}$ and $\Psi^{\infty}$ depend on $(\ell,m,\omega)$.

A sample of results for the case of a point mass on a circular orbit of radius $r_0 = p M$ in the equatorial plane of Kerr spacetime are shown in Fig.~\ref{fig:amplitudes}, and given as a data table in the supplemental material to this paper. These agree to all significant digits with values obtained using the code described in Ref.~\cite{Dolan:2023enf}. Finally, it is clear from the plot that we can read off the power-law behaviour of these asymptotic amplitudes (corresponding to the leading term in a post-Newtonian expansion). Doing so, we find
\begingroup
\allowdisplaybreaks
\begin{IEEEeqnarray*}{rClrCl}
    \psi_4^{\cH} &\approx& -\frac34 i \sqrt{\frac{\pi}{5}} p^{-3/2}, \quad&
    \psi_4^{\infty} &\approx& -8 i \sqrt{\frac{\pi}{5}} p^{-5/2},\\
    \phi_2^{\cH} &\approx& \frac15 \sqrt{\frac{\pi}{5}} p^{0}, \quad&
    \phi_2^{\infty} &\approx& -\frac{16}{3} i \sqrt{\frac{\pi}{5}} p^{1/2},\\
    \chi^{\cH} &\approx& -\frac{6}{25} \sqrt{\frac{2\pi}{15}} p^{0}, \quad&
    \chi^{\infty} &\approx& \frac{11}{4} \sqrt{\frac{2\pi}{15}} p^{2},\\
    h^{\cH} &\approx& \frac{32}{25} \sqrt{\frac{2\pi}{15}} p^{-3}, \quad&
    h^{\infty} &\approx& -8 \sqrt{\frac{2\pi}{15}} p^{-1},\\
    \phi_0^{\cH} &\approx& \frac{32}{125} i \sqrt{\frac{\pi}{5}} p^{0}, \quad&
    \phi_0^{\infty} &\approx& 4 i \sqrt{\frac{\pi}{5}} p^{7/2},\\
    \psi_0^{\cH} &\approx& -\frac{2496}{625} \sqrt{\frac{\pi}{5}} p^{-3/2}, \quad&
    \psi_0^{\infty} &\approx&  -3 i \sqrt{\frac{\pi}{5}} p^{7/2} .
\end{IEEEeqnarray*}
\endgroup

\section{Conclusions}
\label{sec:Conclusion}

We have developed a formalism for solving the Lorenz-gauge linearised Einstein equation in Kerr spacetime. The formalism involves solving only (decoupled, separable) Teukolsky equations and reconstructing the metric perturbation by applying differential operators to the solutions. It builds on previous work for homogeneous solutions \cite{Dolan:2021ijg}, extending those results to the case where there is a non-zero source. Unlike the method developed in Ref.~\cite{Dolan:2023enf} it does not rely a special property of the solution, so this new approach is expected to be much more widely applicable.

The main limitation of the prescription is that, because it determines the \emph{time derivative} of the metric perturbation, it cannot be used to obtain static (zero-frequency) solutions. This is a common issue, also encountered by other approaches \cite{Dolan:2023enf,Berndtson:2007gsc}. As was found in those cases, we anticipate it will be possible to use a separate treatment by specialising to the zero-frequency case. Indeed,
in addition to the AAB solution given in Eq.~\eqref{eq:AAB} there is another ``symmetric'' metric perturbation,
\begin{equation}
\label{eq:AAB-plus}
  h^+_{\mu \nu} = \tfrac43 (\cS_4^\dag \zeta^4 \Psi_0 + \cS_0^\dag \zeta^4 \Psi_4)_{\mu\nu} 
\end{equation}
which satisfies the homogeneous linearised Einstein equation
\begin{equation}
  (\cE h^+)_{\mu\nu} = 0
\end{equation}
and where the potentials satisfy ``circularity relations''
\begin{subequations}
\label{eq:circularity-plus}
\begin{align}
  \cT_0 h^+ &= \edth^4 \bar{\zeta}^4 \bar{\Psi}_0,\\
  \cT_4 h^+ &= \edth'^4 \bar{\zeta}^4 \bar{\Psi}_4.
\end{align}
\end{subequations}
This has the advantage of also being valid in the zero-frequency case. However, a sourced version (i.e. a corrector tensor equivalent to $\cN T$) is not yet known in this case and there is also additional subtlety related to the complex conjugates of the potentials: the metric perturbation in Eq.~\ref{eq:AAB-plus} is complex, and taking the real part then introduces extra terms into the circularity relations in Eq.~\eqref{eq:circularity-plus}. These issues will be addressed in a future work.
    
\begin{acknowledgments}
\textit{Acknowledgements.---}
We thank Stefan Hollands and Vahid Toomani for helpful discussions, and we thank Kevin Cunningham for finding typos in an earlier draft of this paper. This work makes use of the Black Hole Perturbation Toolkit. Many of the calculations in this work were enabled by the \textsc{xAct} \cite{xTensor,xTensorOnline} tensor algebra package for \textsc{Mathematica}. CK acknowledges support from Science Foundation Ireland under Grant number 21/PATH-S/9610. SD~acknowledges financial support from the Science and Technology Facilities Council (STFC) under Grant No.~ST/T001038/1 and Grant No.~ST/X000621/1. 
\end{acknowledgments}

\appendix

\section{Irreducible decompositions, null tetrad projections and symmetries of Kerr spacetime}
\label{sec:irreducible-decompositions}

In this work we have considered perturbations of spacetimes in the Kerr-NUT class (i.e. non-accelerating Petrov type-D). These perturbations are quite naturally expressed using the spinor formalism, as it efficiently handles simplifications that arise from the use of irreducible decompositions of tensors. In this Appendix, we provide a review some concepts tensorial expressions for the key results of interest to this paper and direct the reader to Refs.~\cite{Aksteiner:2016pjt,Aksteiner:2014zyp,Penrose:1985bww} for a more thorough exposition in the language of spinors.

\subsection{Null tetrads}

We introduce an orthonormal basis of null vectors, $\{l^\alpha, n^\alpha, m^\alpha, \mb^\alpha\}$ where $l^{\mu}$ and $n^{\nu}$ are real and are aligned with the principal null directions, $m^{\mu}$ is a complex and $\mb^{\nu}$ is its complex conjugate.  The tetrad satisfies the orthonormality conditions $l^\mu n_\mu = -1$ and $m^\mu \mb_\mu = 1$, with all other inner products being zero. In this basis the spacetime metric is
\begin{equation}
g^{\mu \nu} = - 2 \, l^{(\mu} n^{\nu)} + 2 \, m^{(\mu} \mb^{\nu)}.
\end{equation}

\subsection{Geroch-Held-Penrose derivatives}
Using the null tetrad, we next define the Geroch-Held-Penrose (GHP) directional derivatives \cite{Geroch:1973am} (see Sec.~4.1.1 of Ref.~\cite{Pound:2021qin} for a review):
\begin{alignat}{3}
\th   &:= (l^\alpha \nabla_\alpha - p \epsilon - q \bar{\epsilon}), & \quad
\th'  &:= (n^\alpha \nabla_\alpha + p \epsilon' + q \bar{\epsilon}'), \nonumber \\
\edth  &:= (m^\alpha \nabla_\alpha - p \beta + q \bar{\beta}'),& \quad
\edth' &:= (\bar{m}^\alpha \nabla_\alpha + p \beta' - q\bar{\beta}),
\end{alignat}
where
\begin{subequations}
\begin{align}
  \beta &= \frac{1}{2} (m^\mu \mb^\nu \nabla_\mu m_\nu-m^\mu n^\nu \nabla_\mu l_\nu), \\
  \epsilon &= \frac{1}{2} (l^\mu \mb^\nu \nabla_\mu m_\nu-l^\mu n^\nu \nabla_\mu l_\nu),
\end{align}
\end{subequations}
along with their primed variants $\beta'$ and $\epsilon'$ that are obtained by interchanging the tetrad vectors $l^\alpha \leftrightarrow n^\alpha$ and $m^\alpha \leftrightarrow \bar{m}^\alpha$. Here $\{p,q\}$ represents the GHP type of the object on which the derivatives are acting; they are related to the spin-weight $s=(p-q)/2$ and boost-weight $b=(p+q)/2$. We also introduce the remaining 8 {\it spin coefficients}, defined to be the directional derivatives of the tetrad vectors:
\begin{alignat}{2}
  \kappa = - l^\mu m^\nu \nabla_\mu l_\nu, & \quad \sigma = - m^\mu m^\nu \nabla_\mu l_\nu, &\nonumber \\
  \rho = -\mb^\mu m^\nu \nabla_\mu l_\nu, & \quad \tau = - n^\mu m^\nu \nabla_\mu l_\nu,
\end{alignat}
along with their primed variants, $\kappa'$, $\sigma'$, $\rho'$ and $\tau'$ (for a tetrad aligned to the principal null directions of Kerr spacetime we have $\kappa = \kappa' = \sigma = \sigma' = 0$). Finally, the GHP derivative operators have adjoints given by
\begin{equation}
\cD^\dag = - (\zeta\bar{\zeta})^{-1} \cD (\zeta\bar{\zeta}), \quad \cD \in\{\th, \th', \edth, \edth'\},
\end{equation}
where $\zeta$ is the Killing spinor coefficient (see Sec.~\ref{sec:symmetries}).

\subsection{Tetrad projections and self-dual decompositions}

Any antisymmetric rank-2 tensor (i.e.~a two-form) $F_{\alpha\beta} = F_{[\alpha\beta]}$ can be projected onto the null tetrad,
\begin{align}
  &F_{\alpha\beta} = 2 \Big[
      (\Phi_1+\bar{\Phi}_1) n_{[\alpha} l_{\beta]}  
    + (\Phi_1-\bar{\Phi}_1) m_{[\alpha} \mb_{\beta]} \nonumber \\
   &\,\, + \Phi_0 \mb_{[\alpha} n_{\beta]}
        + \bar{\Phi}_0 m_{[\alpha} n_{\beta]}
        + \Phi_2 l_{[\alpha} m_{\beta]} 
        + \bar{\Phi}_2 l_{[\alpha} \mb_{\beta]}
     \Big].
\end{align}
Such tensors can also be decomposed into self-dual and anti-self-dual parts, $F_{\alpha \beta} = \cF_{\alpha\beta} + \bar{\cF}_{\alpha\beta}$, where
\begin{subequations}
\begin{align}
    \cF_{\alpha\beta} &= \frac12(F_{\alpha\beta} - i\,{}^\star F_{\alpha\beta}),\\
    \bar{\cF}_{\alpha\beta} &= \frac12(F_{\alpha\beta} + i\,{}^\star F_{\alpha\beta}),
\end{align}
\end{subequations}
where ${}^\star F_{\alpha\beta} = \frac{1}{2} \epsilon_{\alpha\beta}{}^{\gamma\delta} F_{\gamma\delta}$ is the Hodge dual of $F_{\alpha\beta}$, and where the (anti-)self-dual property means that ${}^\star \cF_{\alpha\beta} = i \cF_{\alpha\beta}$ and ${}^\star \bar{\cF}_{\alpha\beta} = -i \bar{\cF}_{\alpha\beta}$.  The self-dual part only has components $\Phi_0$, $\Phi_1$ and $\Phi_2$, while the anti-self-dual part only has components $\bar{\Phi}_0$, $\bar{\Phi}_1$ and $\bar{\Phi}_2$.

Similarly, the 10 independent components of the Weyl tensor can be represented by 5 complex Weyl scalars
\begin{gather}
  \psi_0 = C_{l m l m},\quad
  \psi_1 = C_{l n l m},\quad
  \psi_2 = C_{l m \mb n}, \nonumber \\
  \psi_3 = C_{l n \mb n},\quad
  \psi_4 = C_{n\mb n \mb}.
\end{gather}
where $C_{l m l m} \equiv C_{\alpha \beta \gamma \delta} l^\alpha m^\beta l^\gamma m^\delta$, etc. 
The Weyl tensor can also be decomposed into self-dual and anti-self-dual parts, $C_{\alpha \beta\gamma \delta} = \cC_{\alpha \beta\gamma \delta} + \bar{\cC}_{\alpha \beta\gamma \delta}$, where
\begin{subequations}
\begin{align}
    \cC_{\alpha \beta\gamma \delta} &= \frac12(C_{\alpha \beta\gamma \delta} - i\,{}^\star C_{\alpha \beta\gamma \delta}),\\
    \bar{\cC}_{\alpha \beta\gamma \delta} &= \frac12(C_{\alpha \beta\gamma \delta} + i\,{}^\star C_{\alpha \beta\gamma \delta}),
\end{align}
\end{subequations}
and ${}^\star C_{\alpha \beta\gamma \delta} = \frac{1}{2} \epsilon_{\alpha\beta}{}^{\mu\nu} C_{\mu \nu\gamma \delta}$  is the Hodge dual of $C_{\alpha \beta\gamma \delta}$ (N.B.~the left and right duals of the Weyl tensor are identical). The (anti-)self-dual property means that ${}^\star \cC_{\alpha \beta\gamma \delta} = i \cC_{\alpha \beta\gamma \delta}$ and ${}^\star \bar{\cC}_{\alpha \beta\gamma \delta} = -i \bar{\cC}_{\alpha \beta\gamma \delta}$. The self-dual part only has components $\psi_0$, $\psi_1$, $\psi_2$, $\psi_3$ and $\psi_4$, while the anti-self-dual part only has components $\bar{\psi}_0$, $\bar{\psi}_1$, $\bar{\psi}_2$, $\bar{\psi}_3$ and $\bar{\psi}_4$.

\subsection{Irreducible decompositions and covariant derivatives}

The irreducible decomposition of a tensor reduces it to a sum of trace, symmetric-trace-free, and anti-symmetric pieces. For example, a rank-2 tensor can be decomposed as $W_{\alpha \beta} = W_{[\alpha \beta]} + [W_{(\alpha \beta)} - \frac14 g_{\alpha \beta} W] +  \frac14 g_{\alpha \beta} W$. The antisymmetric part can be further decomposed into self-dual and anti-self-dual pieces, $W_{[\alpha \beta]} = \cW_{[\alpha \beta]} + \bar{\cW}_{[\alpha \beta]}$. Similarly, the Riemann tensor can be decomposed into Weyl, trace-free Ricci and trace pieces,
$R_{\alpha \beta \gamma \delta} = C_{\alpha\beta\gamma\delta} + g_{\alpha[\gamma} S_{|\beta|\delta]} + g_{\beta[\delta} S_{|\alpha|\gamma]}+ \frac16 g_{\alpha[\gamma}g_{|\beta|\delta]} R$,
where $S_{\alpha\beta} = R_{\alpha\beta} - \frac14 g_{\alpha\beta} R$ is the trace-free Ricci tensor. The Weyl tensor can be further decomposed into self-dual and anti-self-dual pieces. These irreducible decompositions are naturally represented using their correspondence to symmetric spinors.

The covariant derivative $\nabla_\alpha = \nabla_{AA'}$ of a symmetric spinor can be decomposed into four irreducible parts: the divergence $\Div$, curl $\Curl$, curl-dagger $\CurlDag$, and twistor $\Twist$ operators. For example, the derivative of a vector can be decomposed as
\begin{equation}
    \nabla_\alpha \xi_\beta = \frac12 (\Curl\xi)_{\alpha \beta} + \frac12 (\CurlDag\xi)_{\alpha \beta} + \frac14 g_{\alpha \beta} (\Div\xi) + (\Twist^\dag\xi)_{\alpha \beta}
\end{equation}
where
\begin{subequations}
\begin{align}
  (\Curl\xi)_{\alpha \beta} &= (1-i\,{}^\star)\nabla_{[\alpha}\xi_{\beta]}\\
  (\CurlDag\xi)_{\alpha \beta} &=(1+i\,{}^\star)\nabla_{[\alpha}\xi_{\beta]} \\
  (\Div\xi) &= \nabla^{\alpha}\xi_{\alpha}\\
  (\Twist^\dag\xi)_{\alpha \beta} &= \nabla_{(\alpha}\xi_{\beta)} - \frac14 g_{\alpha \beta} \nabla^{\alpha} \xi_{\alpha}
\end{align}
\end{subequations}

We are particularly interested here in the curl of a symmetric, trace-free rank-2 tensor $S_{\alpha \beta}$, which in tensor form is
\begin{align}
(\Curl S)_{\alpha \beta \gamma} &= \frac12\big[(\nabla S)^\TF_{[\alpha \beta] \gamma} - \frac12 i\, \epsilon_{\alpha \beta}{}^{\delta \epsilon} (\nabla S)^\TF_{\delta \epsilon \gamma} \nonumber \\
&\qquad+(\nabla S)^\TF_{[\alpha \gamma] \beta} - \frac12 i\, \epsilon_{\alpha \gamma}{}^{\delta \epsilon} (\nabla S)^\TF_{\delta \epsilon \beta} \big]
\end{align}
where
\begin{align}
(\nabla S)^\TF_{\alpha \beta \gamma} &= \nabla_{\alpha} S_{\beta \gamma}  - \frac13 (g_{\alpha \beta} \nabla^\delta S_{\delta \gamma} + g_{\alpha \gamma} \nabla^\delta S_{\beta \delta}).
\end{align}
We are also interested in the curl-dagger of this rank-3 tensor, which is given by
\begin{equation}
(\CurlDag \Curl S)_{\alpha \beta} = \frac12 i \epsilon_{(\alpha|}{}^{\gamma \delta \epsilon} \nabla_{\gamma} (\Curl S)_{\delta \epsilon |\beta)}.
\end{equation}

\subsection{Symmetries, Killing spinors and Killing-Yano tensors}
\label{sec:symmetries}

The Kerr--NUT spacetime admits a valence $(2,0)$ Killing spinor satisfying
\begin{equation}
 \nabla_{A'(A} \kappa_{BC)}=0.
\end{equation}
The Killing spinor also satisfies the integrability condition
\begin{equation}
  \Psi_{(ABC}{}^D \kappa_{DE)} = 0
\end{equation}
and the tensor wave equation
\begin{equation}
  \Box \kappa_{AB} = \Psi_{ABCD} \kappa^{CD},
\end{equation}
where $\Psi_{ABCD}$ is the Weyl spinor.
In terms of a spinor dyad $\{o_{A}, \iota_{B}\}$ the Killing spinor is $\kappa_{AB} = -2 \zeta o_{(A} \iota_{B)}$, where
$\zeta \propto \psi_2^{-1/3}$
is the \emph{Killing spinor coefficient}.

The Killing spinor is equivalent to the self-dual 2-form
\begin{equation}
  \kappa_{\alpha\beta} = \frac{1}{2}\Big(f_{\alpha\beta} - i {}^\star f_{\alpha\beta}\Big) = \zeta (m_{[\alpha} \mb_{\beta]}-l_{[\alpha} n_{\beta]}) \equiv \zeta \tilde{\kappa}_{\alpha \beta}.
\end{equation}
Here, $f_{\alpha\beta}$ is a conformal Killing-Yano tensor satisfying
\begin{equation}
    f_{\alpha \beta;\gamma} = g_{\beta\gamma} \TKV_\alpha - g_{\alpha \beta} \TKV_\gamma.
\end{equation}
where 
\begin{equation}
  \TKV_\alpha = \frac13 \nabla^{\beta}f_{\alpha \beta}
\end{equation}
is the time-translation Killing vector.
As a consequence, when differentiating expressions involving the conformal Killing-Yano tensor we will often encounter time derivatives represented by the Lie derivative along this time-translation Killing vector, $\LieT$. This commutes with all other operators, and in Boyer-Lindquist coordinates it is simply a partial derivative with respect to coordinate time $t$.

Finally, the Hodge dual of the conformal Killing-Yano tensor, ${}^\star f_{\alpha\beta} = \tfrac12 \epsilon_{\alpha \beta}{}^{\mu \nu} f_{\mu \nu}$, is a Killing-Yano tensor satisfying
\begin{equation}
    {}^\star f_{\alpha (\beta;\gamma)} = 0.
\end{equation}

\subsection{Spin decomposition}

The tensor $\tilde{\kappa}_{\alpha\beta}$ can be used to define spin-raising ($\cK^0$), sign-flipping ($\cK^1)$ and spin-lowering ($\cK^2$) operators. These can, in turn, be used to define spin-projection operators $\cP^i$, which pick out spin-$i$ components and set all other components to zero. Here, we are especially interested in the spin-2 projected and sign-flipped self-dual Weyl tensor (i.e.~the Weyl tensor with only self-dual, maximum spin-weight components and with the sign of the negative spin-weight components flipped), which is given by
\begin{align}
  \cC^-_{\alpha \beta \gamma \delta} &\equiv (\cK^1 \cP^2 \cC)_{\alpha \beta \gamma \delta}\nonumber \\
  &= -2 \tilde{\kappa}_\alpha{}^{\mu}\Big[ \cC_{\mu \beta \gamma \delta}+ (\tilde{\kappa}^{\rho \sigma} \cC_{\rho \sigma \mu \beta} \tilde{\kappa}_{\gamma \delta} + \tilde{\kappa}^{\rho \sigma} \cC_{\rho \sigma \gamma \delta} \tilde{\kappa}_{\mu \beta})\Big] \nonumber \\
  & = 4 (\psi_0 \,  n_{[\alpha} \mb_{\beta]} n_{[\gamma} \mb_{\delta]} - \psi_4 \,  l_{[\alpha} m_{\beta]} l_{[\gamma} m_{\delta]}).
\end{align}

\subsection{Kerr spacetime}
In Kerr-NUT spacetimes $\psi_2$ is the only non-zero Weyl scalar, and in the Kerr case it is given by
\begin{equation}
\psi_2^{\rm Kerr} = -\frac{M}{\zeta^3}
\end{equation}
where
\begin{equation}
    \zeta = r - i a \cos \theta
\end{equation}
is the Killing spinor coefficient in Kerr spacetime.

\begin{widetext}
\section{Operators in GHP form}
\label{sec:operators}

In this Appendix we give the operators appearing in this paper in GHP form. These operators act on symmetric tensors, $h_{\alpha \beta}$ or $T_{\alpha \beta}$, on vectors $j_\alpha$ or $\xi_\alpha$, or on scalars: $\Psi_0$ of GHP weight $\{4,0\}$, $\Psi_4$ of GHP weight $\{-4,0\}$, $\Phi_0$ of GHP weight $\{2,0\}$, $\Phi_2$ of GHP weight $\{-2,0\}$, or $\Phi$ of GHP weight $\{0,0\}$. The adjoint operators, denoted with ${}^\dag$, are defined such that
\begin{equation*}
  X^{\alpha_1 \cdots \alpha_m} (\cD Y)_{\alpha_1 \cdots \alpha_m} - (\cD^\dag X)^{\beta_1 \cdots \beta_n} Y_{\beta_1 \cdots \beta_n} = \nabla_\alpha v^a
\end{equation*}
for some vector $v^\alpha$ and where the GHP weights of $X$ and $Y$ are such that the products $X (\cD Y)$ and $(\cD^\dag X) Y$ are GHP weight $\{0,0\}$.

These operators are also given electronically as a Mathematica notebook in the supplemental material to this paper.

\subsection{Lie derivative along the timelike Killing vector}
The Lie derivative along the time-translation Killing vector $T^\alpha$ acting on an object of GHP weight $\{p,q\}$ is
\begin{equation}
  \LieT = -\zeta \big( - \rho' \th + \rho \th' + \tau' \edth - \tau \edth') - \frac{p}{2} \zeta \Psi_2 - \frac{q}{2} \bar{\zeta} \bar{\Psi}_2.
\end{equation}

\subsection{Decoupling operators}
The decoupling operators for gravitational perturbations are given by
\begin{subequations}
\label{eq:S}
\begin{align}
\mathcal{S}_0 T &= 
  \frac12 (\edth-\bar{\tau}'-4\tau) 
    \big[(\th-2\bar{\rho}) T_{(lm)} -(\edth-\bar{\tau}') T_{ll}  \big]
  + \frac12(\th-4\rho-\bar{\rho})
    \big[(\edth-2\bar{\tau}') T_{(lm)} - (\th-\bar{\rho}) T_{mm}  \big], \\
\mathcal{S}_4 T &= 
  \frac12(\edth'-\bar{\tau}-4\tau') 
    \big[(\th'-2\bar{\rho}') T_{(n\mb)} -(\edth'-\bar{\tau}) T_{nn}  \big]
  + \frac12(\th'-4\rho'-\bar{\rho}')
    \big[(\edth'-2\bar{\tau}) T_{(n\mb)} - (\th'-\bar{\rho}') T_{\mb\mb}  \big],
\end{align}
and their adjoints are given by
\label{eq:Sdag}
\begin{align}
\label{eq:Sdag0}
(\mathcal{S}_0^\dag \Psi_4)_{\alpha \beta} &= \big[- \frac12 l_\alpha l_\beta (\edth-\tau)(\edth+3\tau) - \frac12 m_{\alpha}m_\beta (\th-\rho)(\th + 3\rho) \nonumber \\
  & \qquad\qquad + \frac12 l_{(\alpha}m_{\beta)} \big((\th-\rho+\bar{\rho})(\edth+3\tau) +(\edth-\tau+\bar{\tau}')(\th+3\rho)\big)\big] \Psi_4, \\
\label{eq:Sdag4}
(\mathcal{S}_4^\dag \Psi_0)_{\alpha \beta} &= \big[- \frac12 n_\alpha n_\beta (\edth'-\tau')(\edth'+3\tau')  - \frac12 \mb_{\alpha}\mb_\beta (\th'-\rho')(\th' + 3\rho')
  \nonumber \\
  & \qquad\qquad + \frac12 n_{(\alpha}\bar{m}_{\beta)} \big((\th'-\rho'+\bar{\rho}')(\edth'+3\tau') +(\edth'-\tau'+\bar{\tau})(\th'+3\rho')\big) \big]\Psi_0.
\end{align}
For spin-weight $s=\pm1$ fields they are given by
\begin{align}
  \mathcal{S}_0 j &= \frac12 \big[(\edth-2\tau-\bar{\tau}') j_l-(\th - 2\rho - \bar{\rho}) j_m\big], \label{eq:S0} \\
  \mathcal{S}_2 j &= \frac12 \big[-(\edth'-2\tau' - \bar{\tau}) j_n + (\th' - 2 \rho' - \bar{\rho}') j_\mb\big], \label{eq:S2} 
\end{align}
and their adjoints are given by
\begin{align}
  \label{eq:S0dag}
  (\mathcal{S}^\dag_0 \Phi_2)_\alpha &= \frac12 \big[- l_\alpha (\edth+\tau) + m_\alpha (\th + \rho) \big]\Phi_2, \\
  \label{eq:S2dag}
  (\mathcal{S}^\dag_2 \Phi_0)_\alpha &= \frac12 \big[n_\mu (\edth'+ \tau') - \mb_\mu (\th' + \rho') \big]\Phi_0.
\end{align}
\end{subequations}

\subsection{Linearised Einstein operator}

Five of the components of the linearised Einstein operator are given by symmetries,
\begin{subequations}
\begin{align}
    (\cE h)_{nn}&=\overline{(\cE h)}'_{ll},\\
    (\cE h)_{l\mb}&=\overline{(\cE h)}_{lm},\\
    (\cE h)_{nm}&=\overline{(\cE h)}'_{lm},\\
    (\cE h)_{n\mb}&=(\cE h)'_{lm},\\
    (\cE h)_{\mb\mb}&=\overline{(\cE h)}_{mm}.
\end{align}
The other five are given by
\label{eq:linearisedEE-tetrad}
\begingroup
\allowdisplaybreaks
\begin{align}
(\cE h)_{ll} &=
 -\big[(\edth'-\tau')(\edth-\bar{\tau}') + \rho(\th'+\rho'-\bar{\rho}') -(\th-\rho)\rho' + \psi_2\big]h_{ll}
 - \big[-(\rho+\bar{\rho})(\th+\rho+\bar{\rho}) +4\rho\bar{\rho}\big]h_{ln} \nonumber\\
&\quad
 - \big[-(\th-3\bar{\rho})(\edth'-\tau'+\bar{\tau}) + \bar{\tau}\th-\bar{\rho}\edth'\big]h_{lm} 
 - \big[-(\th-3\rho)(\edth+\tau-\bar{\tau}') + \tau\th-\rho\edth\big]h_{l\mb} \nonumber\\
&\quad
 - \big[\th(\th-\rho-\bar{\rho})+2\rho\bar{\rho}\big]h_{m\mb},
\\
(\cE h)_{ln} &=
    -\frac12\big[\rho'(\th'-\rho') + \bar{\rho}'(\th'-\bar{\rho}')\big]h_{ll} 
	- \frac12\big[\rho(\th-\rho) + \bar{\rho}(\th-\bar{\rho})\big]h_{nn}
	- \frac12\big[-(\edth'+\tau'+\bar{\tau})(\edth-\tau-\bar{\tau}') \nonumber\\
&\qquad 
    - (\edth'\edth+3\tau\tau'+3\bar{\tau}\bar{\tau}') + 2(\bar{\tau}+\tau')\edth 
		+ (\th-2\bar{\rho})\rho' +(\th'-2\rho')\bar{\rho} - \bar{\rho}'(\th+\rho) 
		- \rho(\th'+\bar{\rho}') -\psi_2 -{\bar\psi}_2\big]h_{ln} \nonumber\\
&\quad
  - \frac12\big[(\th'-2\bar{\rho}')(\edth'-\tau') + \bar{\tau}(\th'+\rho'+\bar{\rho}')
    -\tau'(\th'-\rho')-(2\edth'-\bar{\tau})\bar{\rho}'\big]h_{lm}
	- \frac12\big[(\th'-2\rho')(\edth-\bar{\tau}') \nonumber\\
&\qquad 
    + \tau(\th'+\bar{\rho}'+\rho') -\bar{\tau}'(\th'-\bar{\rho}') -(2\edth-\tau)\rho'\big]h_{l\mb} 
	- \frac12\big[(\th-2\rho)(\edth'-\bar{\tau}) + (\tau'+\bar{\tau})(\th+\bar{\rho}) \nonumber\\
&\qquad 
    -2(\edth'-\tau')\rho -2\bar{\tau}\th\big]h_{nm}
	- \frac12\big[(\th-2\bar{\rho})(\edth-\tau) + (\bar{\tau}'+\tau)(\th+\rho)
	  -2(\edth-\bar{\tau}')\bar{\rho}-2\tau\th\big]h_{n\mb} \nonumber\\
&\quad
  - \frac12\big[-(\edth'-\bar{\tau})(\edth'-\tau') + \bar{\tau}(\bar{\tau}-\tau')\big]h_{mm} 
  - \frac12\big[-(\edth-\tau)(\edth-\bar{\tau}') + \tau(\tau-\bar{\tau}')\big]h_{\mb\mb} \nonumber\\
&\quad
  - \frac12\big[(\edth'+\tau'-\bar{\tau})(\edth-\tau+\bar{\tau}') -2\th'\th 
    +(\edth'\edth-\tau\tau'-\bar{\tau}\bar{\tau}'+\tau\bar{\tau}) - (\psi_2+{\bar\psi}_2) \nonumber\\
&\qquad
    +(\th'-2\rho')\bar{\rho} + (\th-2\bar{\rho})\rho' +\rho(3\th'-2\bar{\rho}')
		+\bar{\rho}'(3\th-2\rho)+ 2\rho\bar{\rho}' +2(\edth'\tau)-\tau\bar{\tau}\big]h_{m\mb},
\\
(\cE h)_{m\mb} &= 
  - \frac12\big[\th'(\th'-\rho'-\bar{\rho}') + 2\rho'\bar{\rho}'\big]h_{ll}
  -\frac12\big[\th(\th-\rho-\bar{\rho}) + 2\rho\bar{\rho}\big]h_{nn}
	-\frac12\big[-(\th'+\rho'-\bar{\rho}')(\th-\rho+\bar{\rho}) \nonumber\\
&\qquad
     -\th'(\th+\rho)+\rho(\th'+\rho'-\bar{\rho}') -\bar{\psi}_2 +(\edth'-\bar{\tau})(\edth-\tau-\bar{\tau}') 
    + \edth'\edth - (\edth-2\bar{\tau}')\tau' - \bar{\tau}(2\edth+\bar{\tau}')  \nonumber\\
&\qquad
		-2\tau(\edth'-\bar{\tau})+2\tau'\bar{\tau}'+\bar{\rho}\bar{\rho}'\big]h_{ln}
  -\frac12\big[-(\th'-2\rho')(\edth'-2\bar{\tau}) 
	  + \bar{\tau}(\th'+2\rho'-2\bar{\rho}') -2\tau'\bar{\rho}'\big]h_{lm} \nonumber\\
&\quad -\frac12\big[-(\th'-2\bar{\rho}')(\edth-2\tau) 
    + \tau(\th'+2\bar{\rho}'-2\rho') -2\bar{\tau}'\rho'\big]h_{l\mb}
	-\frac12\big[-\bar{\tau}(\edth'-\bar{\tau})-\tau'(\edth'-\tau')\big]h_{mm} \nonumber\\
&\quad -\frac12\big[-(\th-2\bar{\rho})(\edth'-2\tau')
    + \tau'(\th-2\rho-2\bar{\rho}) - 2\rho\bar{\tau}+4\tau'\bar{\rho}\big]h_{nm}
	-\frac12\big[-\tau(\edth-\tau)-\bar{\tau}'(\edth-\bar{\tau}')\big]h_{\mb\mb} \nonumber\\
&\quad -\frac12\big[-(\th-2\rho)(\edth-2\bar{\tau}')
    + \bar{\tau}'(\th-2\bar{\rho}-2\rho) -2\bar{\rho}\tau+4\bar{\tau}'\rho\big]h_{n\mb}
	- \frac12\big[2\th'\th -(\th'-\bar{\rho}')\bar{\rho} - (\th-\rho)\rho'  -(\edth'\tau) \nonumber\\
&\qquad -\rho(\th'-\rho'+\bar{\rho}') - \bar{\rho}'(\th+\rho-\bar{\rho}) 
    -(\edth'-2\tau')\bar{\tau}' + \tau(\edth'+2\bar{\tau}) - \tau'(\edth-\bar{\tau}')
		+ \bar{\tau}(\edth+\tau)-\psi_2-\bar{\psi}_2\big]h_{m\mb},
\\
(\cE h)_{lm} &=
   -\frac12\big[(\th'-\rho')(\edth-\bar{\tau}')
    +(\edth-\tau-2\bar{\tau}')\bar{\rho}' -(\edth-\tau)\rho' +\tau(\th'+\rho')\big]h_{ll}
  -\frac12\big[-(\th-\rho+\bar{\rho})(\edth+\tau-\bar{\tau}') \nonumber\\
&\qquad
    -(\edth-3\tau+\bar{\tau}')\bar{\rho} - 2\rho\bar{\tau}'\big]h_{ln}
  -\frac12\big[-(\th'+\bar{\rho}')(\th-2\bar{\rho}) + \rho(\th'+2\rho'-2\bar{\rho}') - 4\rho'\bar{\rho} +3\psi_2 \nonumber\\
&\qquad
    + (\edth'+\bar{\tau})(\edth-2\bar{\tau}') - \tau(\edth'+\tau'-2\bar{\tau}) -\tau'(\tau-4\bar{\tau}')\big]h_{lm}
  - \frac12\big[-\edth(\edth-2\tau) -
2\bar{\tau}'(\tau-\bar{\tau}')\big]h_{l\mb} \nonumber\\
&\quad
  - \frac12\big[\th(\th-2\rho) + 2\bar{\rho}(\rho-\bar{\rho})\big]h_{nm}
	- \frac12\big[-(\th-\bar{\rho})(\edth'-\tau'+\bar{\tau}) + 2\bar{\tau}\bar{\rho}\big]h_{mm} \nonumber\\
&\quad - \frac12\big[(\th+\rho-\bar{\rho})(\edth+\bar{\tau}'-\tau) + 2\bar{\tau}'(\th-2\rho) -
(\edth-\tau-\bar{\tau}')\bar{\rho} +2\rho\tau\big]h_{m\mb},
\\
(\cE h)_{mm} &= 
 -\big[-\edth(\edth-\tau-\bar{\tau}') -2\tau\bar{\tau}'\big]h_{ln} 
  -\big[(\th'-\rho')(\edth-\bar{\tau}') - (\edth-\tau-\bar{\tau}')\rho' + 
	   \tau(\th'+\rho'-\bar{\rho}')-\bar{\tau}'(\th'l+\bar{\rho}')\big]h_{lm} \nonumber\\
&\quad
 -\big[(\th-\bar{\rho})(\edth-\tau) - (\edth-\tau-\bar{\tau}')\bar{\rho} -\tau(\th+\rho)
    + \bar{\tau}'(\th-\rho+\bar{\rho})\big]h_{nm}
  -\big[(\tau+\bar{\tau}')\edth + (\tau-\bar{\tau}')^2\big]h_{m\mb} \nonumber\\
&\quad
 -\big[-(\th'-\rho')(\th-\bar{\rho}) + (\edth-\tau)\tau' -\tau(\edth'+\tau'-\bar{\tau}) + \psi_2\big]h_{mm}.
\end{align}%
\endgroup
The linearised Einstein operator is self-adjoint,
\begin{equation}
    \cE^\dag = \cE.
\end{equation}
\end{subequations}

\subsection{Lorenz gauge vector}
Defining the operator $\cZ$ by $(\cZ h)_\alpha = \nabla^\nu \trev{h}_{\alpha \nu}$, two of its components are given by symmetries,
\begin{subequations}
\begin{align}
    (\cZ h)_n &=(\cZ h)_l',\\
    (\cZ h)_\mb &=(\cZ h)_m',
\end{align} 
and the other two components are given by
\begin{align}
  (\cZ h)_l &=- (\th - \rho - \bar{\rho})\hat{h}_{ln} - (\th'-\rho'-\bar{\rho}')\hat{h}_{ll} + (\edth-\bar{\tau}' - 2 \tau)\hat{h}_{l\mb} + (\edth'-\tau' - 2 \bar{\tau})\hat{h}_{lm} +  (\rho  + \bar{\rho})\hat{h}_{m\mb}, \\
  (\cZ h)_m &=  - (\th-2\rho - \bar{\rho})\hat{h}_{nm} - (\th'-\rho' - 2 \bar{\rho}')\hat{h}_{lm} + (\edth-\bar{\tau}'-\tau)\hat{h}_{m\mb} + (\edth'-\tau' - \bar{\tau})\hat{h}_{mm} - (\tau + \bar{\tau}')\hat{h}_{ln}.f
\end{align}
\end{subequations}

\subsection{Vector wave operator}
Two of the components of the vector wave operator are given by symmetries,
\begin{subequations}
\begin{align}
    (\Box \xi)_{l}&=\overline{(\Box \xi)_{l}'},\\
    (\Box \xi)_{\mb}&=\overline{(\Box \xi)_{m}},
\end{align} 
and the other two components are given by
\begin{align}
  (\Box \xi)_l &= 
    2\big[ (\eth'-\tau')(\edth - \bar{\tau}')-(\th - \rho)(\th' - \bar{\rho}') + \rho \rho' \big] \xi_l 
    +2 \rho \bar{\rho} \xi_n
    +2\big[ \bar{\rho} \edth' - \bar{\tau} \th \big] \xi_m
    +2\big[ \rho \edth - \tau \th \big] \xi_\mb,
    \\
  (\Box \xi)_m &= 
    2\big[ (\edth - \tau)(\edth' - \bar{\tau})-(\th'-\rho')(\th - \bar{\rho}) - \tau \tau' \big] \xi_m 
    -2 \tau \bar{\tau}' \xi_\mb
    +2\big[\bar{\rho}' \edth - \bar{\tau}' \th'\big] \xi_l
    +2\big[ \rho \edth- \tau \th \big] \xi_n.
\end{align}
\end{subequations}
The vector wave operator is self-adjoint.

\subsection{Lorenz gauge scalar}
The divergence of a vector field, $\nabla^\alpha \xi_{\alpha}$, is given by
\begin{equation}
\label{eq:lorenaAGHP}
    \Div \xi = 
    -(\th' - \rho' - \bar{\rho}')\xi_l
    -(\th - \rho - \bar{\rho})\xi_n
    +(\edth' - \tau' - \bar{\tau})\xi_m
    +(\edth - \tau - \bar{\tau}')\xi_\mb.
\end{equation}

\subsection{Teukolsky operators}
The Teukolsky operators for spin-weight $s=\pm2$ fields are given by
\begin{subequations}
\begin{align}
  \cO_0\Psi_0 &= \big[\big(\th - 4 \rho - \bar{\rho}\big)\big(\th'-\rho'\big) - \big(\edth - 4\tau - \bar{\tau}'\big)\big(\edth' - \tau'\big) -3 \psi_2\big]\Psi_0,
  \\
  \cO_4 \Psi_4 &= \big[\big(\th' -4 \rho' - \bar{\rho}'\big)\big(\th-\rho\big) - \big(\edth' -4\tau' - \bar{\tau}\big)\big(\edth - \tau\big) -3 \psi_2\big]\Psi_4,
\end{align}
and their adjoints are given by
\begin{align}
  \cO_0^\dag \Psi_4 &= \zeta^4 \cO_4 \zeta^{-4} \Psi_4, \nonumber \\
  \cO_4^\dag \Psi_0 &= \zeta^4 \cO_0 \zeta^{-4} \Psi_0.
\end{align}
For spin-weight $s=\pm1$ they are given by
\begin{align}
  \cO_0 \Phi_0 &= \big[\big(\th - 2 \rho - \bar{\rho}\big)\big(\th'-\rho'\big) - \big(\edth - 2\tau - \bar{\tau}'\big)\big(\edth' - \tau'\big) \big]\Phi_0,
  \\
  \cO_2 \Phi_2 &= \big[\big(\th' -2 \rho' - \bar{\rho}'\big)\big(\th-\rho\big) - \big(\edth' +2\tau' - \bar{\tau}\big)\big(\edth - \tau\big) \big]\Phi_2,
\end{align}
and their adjoints are given by
\begin{align}
  \cO_0^\dag \Phi_2 &= \zeta^2 \cO_2 \zeta^{-2} \Phi_2, \nonumber \\
  \cO_2^\dag \Phi_0 &= \zeta^2 \cO_0 \zeta^{-2} \Phi_0.
\end{align}
For spin-weight $s=0$ the operator is given by
\begin{align}
  \cO \Phi &= \big[\big(\th - \bar{\rho}\big)\big(\th'-\rho'\big) - \big(\edth - \bar{\tau}'\big)\big(\edth' - \tau'\big) - \psi_2\big]\Phi,
\end{align}
and its adjoint is given by
\begin{align}
  \cO^\dag \Phi &= \cO \Phi.
\end{align}

\end{subequations}

\subsection{Linearised Weyl operators}
The linearised Weyl operators are given by
\begin{subequations}
\begin{align}
  \mathcal{T}_0 h &= - \frac12 \Big[(\edth - \bar{\tau}')(\edth-\bar{\tau}') h_{ll} + (\th-\bar{\rho})(\th-\bar{\rho}) h_{mm} 
     - \big((\th-\bar{\rho})(\edth-2\bar{\tau}')+ (\edth-\bar{\tau}')(\th-2\bar{\rho})\big) h_{(lm)} \Big],\label{eq:T0} \\
  \mathcal{T}_4 h &= - \frac12 \Big[(\edth' - \bar{\tau})(\edth'-\bar{\tau}) h_{nn} + (\th'-\bar{\rho}')(\th'-\bar{\rho}') h_{\mb\mb} 
  - \big((\th'-\bar{\rho}')(\edth'-2\bar{\tau})+ (\edth'-\bar{\tau})(\th'-2\bar{\rho}')\big) h_{(n\mb)} \Big]\label{eq:T4},
\end{align}
and their adjoints are given by
\begin{align}
  (\mathcal{T}_0^\dag \Psi_4)_{\alpha \beta} &= - \frac12 \Big[l_\alpha l_\beta (\edth - \tau)(\edth - \tau) + m_\alpha m_\beta (\th-\rho)(\th-\rho) \nonumber \\
     & \qquad \qquad
  - l_{(\alpha} m_{\beta)} \big((\edth-\tau+\bar{\tau}')(\th-\rho)+ (\th-\rho+\bar{\rho})(\edth-\tau)\big) \Big]\Psi_4,\label{eq:T0dag} \\
  (\mathcal{T}_4^\dag \Psi_0)_{\alpha \beta} &= - \frac12 \Big[n_\alpha n_\beta (\edth' - \tau')(\edth' - \tau') + \mb_\alpha \mb_\beta (\th'-\rho')(\th'- \rho')\nonumber \\
     & \qquad \qquad- n_{(\alpha} \mb_{\beta)} \big((\edth'-\tau'+\bar{\tau})(\th'-\rho') + (\th'-\rho'+\bar{\rho}')(\edth'-\tau')\big)  \Big]\Psi_0.\label{eq:T4dag}
\end{align}
\end{subequations}

\subsection{Maxwell scalar operators}
The operators that give Maxwell scalars in terms of a vector field $\xi_\alpha$ are given by
\begin{subequations}
\begin{align}
  \mathcal{T}_0 \xi &= - (\edth-\bar{\tau}') \xi_l + (\th - \bar{\rho})\xi_m,\label{eq:T0EM} \\
  \mathcal{T}_2 \xi &= (\edth'-\bar{\tau}) \xi_n -(\th' - \bar{\rho}')\xi_{\bar{m}}, \label{eq:T2EM}
\end{align}
and their adjoints are given by
\begin{align}
  (\mathcal{T}^\dag_0 \Phi_2)^\alpha &= l^\alpha (\edth-\tau) \Phi_2 - m^\alpha (\th - \rho) \Phi_2,\label{eq:T0dagEM} \\
  (\mathcal{T}^\dag_2 \Phi_0)^\alpha &= - n^\alpha(\edth'-\tau') \Phi_0 + \mb^\alpha (\th' - \rho')\Phi_0. \label{eq:T2dagEM}
\end{align}
\end{subequations}

\subsection{AAB corrector \texorpdfstring{$\cN$}{N}}
Five of the components of the operator $\cN$ are given by symmetries:
\begin{subequations}
\begin{align}
(\cN T)_{m\mb} &= (\cN T)_{ln},\\
(\cN T)_{nn} &=- (\cN T)_{ll}',\\
(\cN T)_{\mb\mb} &=- (\cN T)_{mm}',\\
(\cN T)_{n\mb} &=- (\cN T)_{lm}',\\
(\cN T)_{nm} &=- (\cN T)_{l\mb}'.
\end{align}
The other five are given by
\begingroup
\allowdisplaybreaks
\begin{align}
 (\cN T)_{ll} &=
   \frac19 \Big[3 (\edth' - \tau') (\edth +  4 \tau - \bar{\tau}') + (\th - \rho) (\th' + 4 \rho' - \bar{\rho}') - 2 \rho \rho' - 9 \psi_2\Big] \zeta^4 T_{ll}
   - \frac{1}{54} \th \zeta^4 \th T 
   + \frac{1}{9} \rho \Big[2 \rho - \bar{\rho}\Big] \zeta^4 \tilde{T} 
   \nonumber \\ & \quad
   + \frac{2}{9} \Big[\tau \th -\rho (\edth + 4 \tau)\Big] \zeta^4 T_{l\mb}
   - \frac{1}{9} \Big[(2 \edth' - \bar{\tau}) (2 \th + 6 \rho - 3 \bar{\rho}) - 8 \rho \tau' - 5 \bar{\rho} \bar{\tau}\Big] \zeta^4 T_{lm}
   + \frac{2}{27} \th \zeta^4 (\Div T)_l,
   \\
 (\cN T)_{mm} &= 
   -\frac{1}{9}\Big[3 (\th'-\rho')(\th+4 \rho- \bar{\rho}) + (\edth- \tau)(\edth'+4  \tau'-\bar{\tau})- 2 \tau\tau'+9\psi_2\Big] \zeta^4 T_{mm}
  -\frac{1}{54}\edth \zeta^4 \edth T
  -\frac{1}{9} \tau \Big[2\tau - \bar{\tau}'\Big] \zeta^4 \tilde{T}
  \nonumber \\ & \quad 
  +\frac{2}{9} \Big[\tau\th-\rho(\edth-4\tau)\Big]\zeta^4 T_{nm}
  +\frac{1}{9} \Big[(2\th'- \bar{\rho}')(2\edth+6 \tau-3 \bar{\tau}')-8 \tau \rho'-5\bar{\rho}'\bar{\tau}'\Big] \zeta^4 T_{lm}
  +\frac{2}{27} \edth \zeta^4 (\Div T)_m,
  \\
 (\cN T)_{lm} &= 
  \frac{1}{9}\Big[(\edth'-4 \tau'+3 \bar{\tau})(\edth+7 \tau-4 \bar{\tau}')-(\th'-4 \rho'+3 \bar{\rho}')(\th+7 \rho-4 \bar{\rho})-14 \psi_2+2 \bar{\psi}_2-26 \rho\rho'+32 \bar{\rho}\rho' -10 \bar{\rho}'\bar{\rho}
  \nonumber \\ & \quad \qquad
  -32 \tau\bar{\tau}+26 \tau\tau'+10 \bar{\tau}\bar{\tau}'\Big] \zeta^4 T_{lm}
  +\frac{1}{18} \Big[\tau\th-\rho\edth\Big]\zeta^4 \tilde{T}
  -\frac{2}{9} \Big[\tau(2\tau- \bar{\tau}')\Big]\zeta^4 T_{l\mb}
  +\frac{2}{9} \Big[ \rho(2 \rho- \bar{\rho})\Big]\zeta^4 T_{nm}
  \nonumber \\ & \quad
  +\frac{1}{18} \Big[(2\th'+\bar{\rho}')(2\edth+6\tau- \bar{\tau}')-8 \tau\rho'-5 \bar{\rho}'\bar{\tau}'\Big]\zeta^4 T_{ll}
  -\frac{1}{18} \Big[(2\edth'+ \bar{\tau})(2\th+6 \rho- \bar{\rho})-8 \rho \tau'-5 \bar{\rho}\bar{\tau}\Big]\zeta^4 T_{mm}
  \nonumber \\ & \quad 
  -\frac{1}{108}\Big[(\edth+\bar{\tau}')\zeta^4 \th +(\th+\bar{\rho})\zeta^4 \edth\Big] T
  +\frac{1}{27} \Big[\th+\bar{\rho}\Big]\zeta^4 (\Div T)_m
  +\frac{1}{27} \Big[\edth+\bar{\tau}'\Big]\zeta^4 (\Div T)_l,
  \\
 (\cN T)_{l\mb} &= 
  -\frac{1}{27}\Big[(\edth'-5 \tau')(\edth+8 \tau-\bar{\tau}')+3 (\th- \rho)(\th'+4 \rho'- \bar{\rho}')- 6 \rho\rho'+18 \tau\tau'\Big]\zeta^4 T_{l\mb}
  -\frac{1}{27}\Big[\th+\rho\Big]\zeta^4 (\Div T)_\mb
  \nonumber \\ & \quad
  -\frac{1}{27} \Big[\edth'+\tau'\Big]\zeta^4 (\Div T)_l
  -\frac{1}{54} \Big[(2\edth'- \tau')(2\th'+9 \rho'- 2\bar{\rho}')-9 \rho'\tau'\Big]\zeta^4 T_{ll}
  -\frac{2}{9} \Big[\rho(2 \rho-\bar{\rho})\Big]\zeta^4 T_{n\mb}
  \nonumber \\ & \quad
  +\frac{1}{108}\Big[(\edth'+\tau')\zeta^4 \th +(\th+\rho)\zeta^4 \edth'\Big]T
  +\frac{1}{108}\Big[(2\edth'+2\tau'-3 \bar{\tau})(2\th+4 \rho- \bar{\rho})-9 \bar{\rho}\bar{\tau}-24 \rho\tau'\Big]\zeta^4 \tilde{T}
  \nonumber \\ & \quad
  +\frac{1}{9} \Big[\rho(\edth+8 \tau)- \tau(\th+4 \rho)\Big]\zeta^4 T_{\mb\mb}
  +\frac{2}{27} \Big[(\edth'+4 \tau'-3 \bar{\tau})(\edth'+\bar{\tau})-3 \bar{\tau}\tau'\Big]\zeta^4 T_{lm},
 \\
 (\cN T)_{ln} &=
   -\frac{1}{27} \zeta^4\Big[(2\rho'-\bar{\rho}')\th-(2\rho-\bar{\rho})\th'-(2\tau'-\bar{\tau})\edth+(2\tau-\bar{\tau}')\edth'\Big]\tilde{T}
   \nonumber \\ & \quad
   +\frac{1}{54} \Big[(\edth'-2 \tau'+ \bar{\tau})(\edth'- \bar{\tau}+6 \tau')\Big]\zeta^4 T_{mm}
   -\frac{1}{54} \Big[(\edth-2 \tau+ \bar{\tau}')(\edth- \bar{\tau}'+6 \tau)\Big]\zeta^4 T_{\mb\mb}
   \nonumber \\ & \quad
   -\frac{1}{54} \Big[(\th'-2 \rho'+ \bar{\rho}')(\th'+6 \rho'- \bar{\rho}')\Big]\zeta^4 T_{ll}
   +\frac{1}{54} \Big[(\th-2 \rho+ \bar{\rho})(\th+6 \rho- \bar{\rho})\Big]\zeta^4 T_{nn}
   \nonumber \\ & \quad
   -\frac{2}{27} \Big[(\edth+ \tau)(\th'+2 \rho')-2\bar{\rho}'\bar{\tau}'-6 \rho'\tau\Big]\zeta^4 T_{l\mb}
   -\frac{2}{27} \Big[(2\rho'-\bar{\rho}')(\edth'+2 \tau')-(2\tau'-\bar{\tau})(\th'+2\rho')\Big]\zeta^4 T_{lm}
   \nonumber \\ & \quad
   +\frac{2}{27} \Big[(\edth'+ \tau')(\th+2 \rho)-2\bar{\rho}\bar{\tau}-6 \rho\tau'\Big]\zeta^4 T_{nm}
   +\frac{2}{27} \Big[(2\rho-\bar{\rho})(\edth+2 \tau)-(2\tau-\bar{\tau}')(\th+2 \rho)\Big]\zeta^4 T_{n\mb}.
\end{align}%
\endgroup
where $T = 2(T_{m \mb} - T_{ln})$ is the trace, $\tilde{T} := 2(T_{m \mb}+T_{ln})$, and $(\Div T)_a = \nabla^\alpha T_{\alpha a}$ is the divergence of $T_{\alpha \beta}$. The operator $\cN$ is anti-self-adjoint,
\begin{equation}
    \cN^\dag = - \cN.
\end{equation}
\end{subequations}

\subsection{AAB vector \texorpdfstring{$\cA$}{A}}
The AAB vector operator $\cA$ itself is not used in the Lorenz-gauge construction, but is included here for completeness. Two of the components of the operator $\cA$ are given by symmetries:
\begin{subequations}
\begin{align}
     (\cA h)_{n} &= -(\cA h)_{l}' \\ 
     (\cA h)_{\mb} &= -(\cA h)_{m}' 
\end{align}
The other two are given by
\begingroup
\allowdisplaybreaks
\begin{align}
 (\cA h)_{l} &= 
  \frac{\zeta^4}{108}\Big\{ \big[
  \th'\th'\th
  +6 \edth'\edth\th'
  -6 \rho \th'\th'
  +2\left( 2 \rho'-\bar{\rho}'\right) \th'\th
  +2\left( \bar{\tau} - 16 \tau' \right) \edth\th'
  +2\left( \tau  - 4 \bar{\tau}'\right) \edth'\th'
  +6\left(3 \rho'- \bar{\rho}'\right) \edth'\edth
  \nonumber \\ & \quad
  +\left(4 \rho'^2(5 \bar{\rho} -3 \rho) +8\rho' (7 \tau \bar{\tau}  - \tau' \tau)+4 \bar{\rho}'( \bar{\tau}' \bar{\tau} +5 \tau \bar{\tau}-4 \rho' \bar{\rho})  +14 \psi_2 (\bar{\rho}' -\rho')-2 \bar{\psi}_2(\bar{\rho}'  +5 \rho')  -12 \rho' \bar{\psi}_2 \bar{\zeta}/\zeta\right)
  \nonumber \\ & \quad
  +2\rho'\left(3 \rho'-2 \bar{\rho}'\right) \th
  +\left(18 \rho' \bar{\rho}-20 \rho' \rho -8 \bar{\rho}' \bar{\rho} +8 \tau' \tau -4 \bar{\tau}' \bar{\tau} +26 \tau \bar{\tau} -5 \psi_2 -5 \bar{\psi}_2 -2 \bar{\psi}_2\bar{\zeta}/\zeta +4 \bar{\psi}_2\bar{\zeta}^2/\zeta^2 \right) \th'
  \nonumber \\ & \quad
  -\left(78 \rho' \tau'+8 \rho' \bar{\tau} +12 \bar{\rho}' \bar{\tau} \right) \edth
  +2\left(9 \rho' \tau +8 \bar{\rho}' \tau \right) \edth'
  \big]h_{ll}
  \nonumber \\ & 
  + \big[
  -2 \th'\th\th
  -8 \edth'\edth\th
  +32 \tau' \edth\th
  -2\left( \rho'- \bar{\rho}'\right) \th\th
  +4\left(2\rho +\bar{\rho} \right) \th'\th
  +4\left(5 \bar{\tau}' -3 \tau \right) \edth'\th
  +4\left( \rho + \bar{\rho} \right) \edth'\edth
  \nonumber \\ & \quad
  +\left(4\rho' (5  \rho^2-7 \bar{\rho}^2) +4\bar{\rho} (5 \bar{\rho}' \bar{\rho} -7 \rho' \rho)  +8\rho \tau(5 \tau' -3 \bar{\tau})  +8\bar{\rho} \bar{\tau}(5 \bar{\tau}'  -3 \tau ) -8\psi_2(7 \rho  +3 \bar{\rho} ) +4 \bar{\rho} \bar{\psi}_2 -12  \rho \bar{\psi}_2 \bar{\zeta}/\zeta\right)
  \nonumber \\ & \quad
  +\left(4\bar{\rho}(5 \bar{\rho}'-3 \rho') -4 \rho \rho'  +16( \tau \tau' + \bar{\tau} \bar{\tau}') -56 \tau \bar{\tau} -6 \psi_2 +\bar{\psi}_2 (8 +6 \bar{\zeta} /\zeta-12  \bar{\zeta}^2/\zeta^2)\right) \th
  +4\left(5 \rho^2 + \bar{\rho}^2 -6 \rho \bar{\rho} \right) \th'
  \nonumber \\ & \quad
  +4\left(7 \rho\tau'  +4 \rho \bar{\tau} +3 \bar{\rho} \bar{\tau} \right) \edth
  +16\left( \bar{\rho} \tau+ \bar{\tau}' \bar{\rho}\right)\edth'
  \big]h_{ln}
  \nonumber \\ &
  + \big[
  -8 \edth'\th'\th
  -6 \edth'\edth'\edth
  +4\left( \bar{\tau}+7 \tau'\right) \th'\th
  -8\left(2 \rho'- \bar{\rho}'\right) \edth'\th
  +4\left(\rho+3 \bar{\rho}\right) \edth'\th'
  +4\left(\bar{\tau}+5 \tau'\right) \edth'\edth
  \nonumber \\ &\quad
  +2\left(7 \bar{\tau}' -4 \tau \right)\edth'\edth'
  +4\left(10 \rho' \tau' +5 \rho' \bar{\tau} +4 \bar{\rho}' \bar{\tau} \right) \th
  +4\left(8 \tau' \rho +7 \rho \bar{\tau} +4 \bar{\rho}\bar{\tau} \right) \th'
  +4\left(13 \tau'^2 +2 \bar{\tau}^2-9 \tau' \bar{\tau} \right) \edth
  \nonumber \\ &\quad
  +\big(4 \bar{\tau}^2(7 \bar{\tau}' -10 \tau) +48 \tau' (\rho' \rho + \tau' \tau) +40 \rho'  \bar{\tau}(\rho +\bar{\rho})  +4 \bar{\tau}(14 \bar{\rho}' \bar{\rho} - 19 \tau' \tau ) +8 \psi_2 (\tau' -4 \bar{\tau}) +4 \bar{\tau} \bar{\psi}_2 (1-5 \bar{\zeta}^2/\zeta^2)
  \nonumber \\ &\qquad
  -4  \tau' \bar{\psi}_2 \bar{\zeta}/\zeta \big)
  +\left(8 \rho' (\bar{\rho} -2 \rho) +28 (\bar{\rho}' \bar{\rho}- \tau \bar{\tau}) +12(2 \bar{\tau} \bar{\tau}'- \tau \tau')
   +2 \psi_2 +6 \bar{\psi}_2 +12  \bar{\psi}_2 \bar{\zeta}/\zeta-20  \bar{\psi}_2 \bar{\zeta}^2/\zeta^2\right) \edth'
  \big]h_{lm}
  \nonumber \\ & 
  + \big[
  4 \edth\th'\th
  +6 \edth'\edth\edth
  +4\left(\tau-2 \bar{\tau}' \right) \th'\th
  -4\left(  \bar{\rho}'-2  \rho'\right) \edth\th
  +4\left(  \bar{\rho}-6  \rho\right) \edth\th'
  +2\left(2 \bar{\tau} -17 \tau' \right) \edth\edth
  +16\left(\tau -\bar{\tau}'\right) \edth'\edth
  \nonumber \\ &\quad
  +( 4\tau^2 (8 \bar{\tau}-5 \tau') +4\bar{\tau}' (2 \bar{\rho}' \bar{\rho}  -  \tau \bar{\tau})-40  \tau(\rho \rho' + \bar{\rho}\bar{\rho}' )  +16 \rho' \bar{\rho} \tau+4\psi_2 (6 \bar{\tau}' +10 \tau) +4 \bar{\psi}_2 (\bar{\tau}' -7\tau) -20 \tau \bar{\psi}_2 \bar{\zeta} / \zeta
  \nonumber \\ & \qquad
  -4 \bar{\tau}' \bar{\psi}_2 \bar{\zeta}^2 /\zeta^2)
  +4\left(3 \rho' \tau +2 \bar{\rho}' \tau \right) \th
  -4\left(4 \rho \tau +5 \bar{\rho} \tau \right) \th'
  +4\left( \bar{\tau}'^2 +5 \tau^2 -6 \bar{\tau}' \tau \right) \edth'
  \nonumber \\ &\quad
  +\left(32 \rho' (\bar{\rho}-\rho) -16 \bar{\rho}' \bar{\rho} +4\tau(17 \bar{\tau} -12 \tau')  -12 \bar{\tau}' \bar{\tau}-36 \psi_2 -10 \bar{\psi}_2 -8  \bar{\psi}_2 \bar{\zeta}/\zeta+6 \bar{\psi}_2 \bar{\zeta}^2 /\zeta^2\right) \edth
  \big]h_{l\mb}
  \nonumber \\ &
  + \big[
  \th\th\th
  -2\left( \rho +\bar{\rho} \right) \th\th
  -2\left(7 \rho^2 +\bar{\rho}^2 -4 \rho \bar{\rho} \right) \th
  +4\left(5 \rho^2 \bar{\rho} -6 \rho^3 -\rho \bar{\rho}^2 \right)
  \big]h_{nn}
  \nonumber \\ &
  + \big[
  10 \edth'\th\th
  -16 \bar{\rho} \edth'\th
  -2\left(15 \tau'+2  \bar{\tau}\right) \th\th
  -4\left(5 \rho^2 +1 \bar{\rho}^2 -2 \rho \bar{\rho} \right) \edth'
  \nonumber \\ & \quad
  -12\left(5 \tau' \rho^2 +2 \rho^2 \bar{\tau} + \rho \bar{\rho} \bar{\tau} \right)
  -8\left(5 \tau' \rho +4 \rho \bar{\tau} +2 \bar{\rho} \bar{\tau} \right) \th
  \big]h_{nm}
  \nonumber \\ &
  + \big[
  -2 \edth\th\th
  + 8 \rho \edth\th
  +2\left(3\bar{\tau}' -2 \tau \right) \th\th
  +8\left( \bar{\tau}' \bar{\rho} + \rho \tau +2 \bar{\rho} \tau \right) \th
  +4\left(5 \rho^2 + \bar{\rho}^2 -4 \rho \bar{\rho} \right) \edth
  \nonumber \\ & \quad
  +12\left( \bar{\tau}' \bar{\rho}^2+4 \rho^2 \tau +2 \bar{\rho}^2 \tau - \rho \bar{\rho} \tau \right)
  \big]h_{n\mb}
  \nonumber \\ & 
  + \big[
  7 \edth'\edth'\th
  -8\left(2\tau'+  \bar{\tau}\right) \edth'\th
  +2\left(\rho-3  \bar{\rho}\right) \edth'\edth'
  -\left(46 \tau'^2 +6 \bar{\tau}^2 -28\tau' \bar{\tau} \right) \th
  -2\left(\tau' \rho +2 \rho \bar{\tau} +4 \bar{\rho} \bar{\tau} \right) \edth'
  \nonumber \\ & \quad
  +4\left(4 \rho \bar{\tau}^2 -12 \tau'^2 \rho -7 \tau' \rho \bar{\tau} \right)
  \big]h_{mm}
  \nonumber \\ & 
  + \big[
  -2 \th'\th\th
  -8 \edth'\edth\th
  +32 \tau' \edth\th
  -2\left(  \rho'- \bar{\rho}'\right) \th\th
  +4\left(2 \rho + \bar{\rho} \right) \th'\th
  +4\left(5 \bar{\tau}' -3 \tau \right) \edth'\th
  +4\left(\rho +\bar{\rho} \right) \edth'\edth
  \nonumber \\ & \quad
  +\left(4\rho \rho' (5 \rho -7 \bar{\rho})+4\bar{\rho}^2(5 \bar{\rho}' -7 \rho')  +4 \rho \tau(10 \tau' -6 \bar{\tau}) +4\bar{\rho} \bar{\tau}(10 \bar{\tau}' -6 \tau ) +8\psi_2(2 \rho -3 \bar{\rho}) +4 \bar{\rho} \bar{\psi}_2 -12 \rho \bar{\psi}_2 \bar{\zeta} /\zeta\right)
  \nonumber \\ & \quad
  +\left(4\bar{\rho}(5\bar{\rho}' -3 \rho') -4 \rho' \rho  +16 (\tau \tau' + \bar{\tau}' \bar{\tau}) -56 \tau \bar{\tau} +42 \psi_2 +8 \bar{\psi}_2 +6 \bar{\psi}_2 \bar{\zeta} /\zeta-12 \bar{\psi}_2 \bar{\zeta}^2 /\zeta^2\right) \th
  \nonumber \\ & \quad
  +4\left(5 \rho^2 +\bar{\rho}^2 -6 \rho \bar{\rho} \right) \th'
  +4\left(7 \tau'\rho +4 \rho \bar{\tau} +3 \bar{\rho} \bar{\tau} \right) \edth
  +16\bar{\rho} \left(\tau +  \bar{\tau}'\right)\edth'
  \big]h_{m\mb}
  \nonumber \\ & 
  + \big[
  \edth\edth\th
  +4\left(\tau -\bar{\tau'}\right) \edth\th
  +2\left(  \bar{\rho}-3 \rho\right) \edth\edth
  +2\left( \bar{\tau}'^2 +3 \tau^2 -4 \bar{\tau}' \tau \right) \th
  -2\left( \bar{\tau}' \bar{\rho} +10 \rho \tau +2 \bar{\rho} \tau \right) \edth
  \nonumber \\ & \quad
  -4\left(6 \rho \tau^2 +2 \bar{\rho} \tau^2 + \bar{\tau}' \bar{\rho} \tau \right)
  \big]h_{\mb\mb} \Big\},
\\
 (\cA h)_{m} &= 
  \frac{\zeta^4}{108}\Big\{\big[
  7 \edth\th'\th'
  -6(2 \tau + \bar{\tau}') \th'\th'
  -2(\rho'+4 \bar{\rho}') \edth\th'
  +4(\bar{\rho}' \tau -7 \rho' \tau -2 \bar{\rho}' \bar{\tau}') \th'
  +\left(20 \bar{\rho}' \rho'-48 \rho'^2-6 \bar{\rho}'^2\right)\edth
  \nonumber \\ & \quad
  + 4\left(4 \bar{\rho}'^2 \tau -18 \tau  \rho'^2-3 \bar{\rho}' \tau  \rho'\right)
  \big]h_{ll}
  \nonumber \\ & 
  + \big[
  -2 \edth'\edth\edth
  -8 \edth\th'\th
  +12(\tau + \bar{\tau}')\th'\th
  +24 \rho' \edth\th
  +12(\bar{\rho}-\rho ) \edth\th'
  +2(\bar{\tau}-\tau') \edth\edth
  +4(2\tau +\bar{\tau}') \edth'\edth
  \nonumber \\ & \quad
  +\left(20 (\bar{\tau} \bar{\tau}'^2+ \tau ^2 \tau')+40 (\rho  \rho' \tau+ \bar{\rho} \bar{\rho}' \bar{\tau}') -24 \bar{\rho} \tau( \rho' - \bar{\rho}') -28 \tau\bar{\tau}(\tau+ \bar{\tau}')-8\psi_2(2 \tau -3 \bar{\tau}')-4 \bar{\tau}' \bar{\psi}_2(1+3 \bar{\zeta}^2/\zeta ^2)\right)
  \nonumber \\ & \quad
  +4(9 \rho'\tau +10 \bar{\rho}' \tau +5 \bar{\rho}' \bar{\tau}') \th
  +4(3 \rho  \tau -7\bar{\rho} \tau +4 \bar{\rho} \bar{\tau}') \th'
  +4\left(5 \tau ^2-6 \bar{\tau}' \tau + \bar{\tau}'^2\right) \edth'
  \nonumber \\ & \quad
  +\left(4 \rho  \rho'-12 \bar{\rho} \rho'+8 \bar{\rho} \bar{\rho}'-4 \tau \bar{\tau}-4 \tau  \tau'+12 \bar{\tau} \bar{\tau}'-42 \psi_2-4 \bar{\psi}_2+6 \bar{\psi}_2 \bar{\zeta}/\zeta-8 \bar{\psi}_2 \bar{\zeta}^2/\zeta^2 \right) \edth
  \big] h_{ln}
  \nonumber \\ & 
  + \big[
  -6 \th'\th'\th
  -8 \edth'\edth\th'
  +4(5 \rho'+ \bar{\rho}') \th'\th
  +2(7 \bar{\rho}-4 \rho ) \th'\th'
  +16( \bar{\tau}- \tau') \edth\th'
  +12( \tau + \bar{\tau}') \edth'\th'
  +4(5 \rho'- \bar{\rho}') \edth'\edth
  \nonumber \\ & \quad
  + \left(4 \rho'^2(16\rho -23 \bar{\rho})+4\bar{\rho} \bar{\rho}'(7 \bar{\rho}'-10 \rho')-56\rho' \tau(   \bar{\tau}-2 \tau')+8\bar{\rho}'\bar{\tau}(5 \tau +7 \bar{\tau}')+4\bar{\rho}' (8 \psi_2- \bar{\psi}_2)-32 \rho' \bar{\psi}_2 \bar{\zeta}/\zeta\right)
  \nonumber \\ & \quad
  +4\left(13 \rho'^2-9 \bar{\rho}' \rho'+2 \bar{\rho}'^2\right)\th
  -4(3 \rho' \bar{\tau}-2 \bar{\rho}' \bar{\tau}-4 \rho'\tau') \edth
  +4(12 \rho' \tau +8 \bar{\rho}' \tau +7 \bar{\rho}' \bar{\tau}') \edth'
  \nonumber \\ & \quad
  -\left(4 \rho  \rho'+36 \bar{\rho} \rho'-24 \bar{\rho} \bar{\rho}'+12 \tau  \bar{\tau}-8 \tau \tau'-16 \bar{\tau} \bar{\tau}'-2 \psi_2+6 \bar{\psi}_2-8 \bar{\psi}_2\bar{\zeta}/\zeta +20 \bar{\psi}_2 \bar{\zeta}^2/\zeta ^2\right) \th'
  \big]h_{lm}
  \nonumber \\ & 
  + \big[
  10 \edth\edth\th'
  -4(5 \tau +4 \bar{\tau}') \edth\th'
  -2(5 \rho'+2 \bar{\rho}') \edth\edth
  -4\left(5 \tau ^2-6 \bar{\tau}' \tau + \bar{\tau}'^2\right) \th'
  -8(5 \rho' \tau +2 \bar{\rho}' \tau +2 \bar{\rho}' \bar{\tau}') \edth
  \nonumber \\ & \quad
  -4\left(15 \rho' \tau ^2+6 \bar{\rho}' \tau ^2+3 \bar{\rho}' \bar{\tau}' \tau \right) 
  \big]h_{l\mb}
  \nonumber \\ & 
  + \big[
   \edth\th\th
  +2(2 \rho - \bar{\rho}) \edth\th
  -6 \tau  \th\th
  -4(5 \rho  \tau -2 \bar{\rho} \tau + \bar{\rho} \bar{\tau}') \th
  +2\left(3 \rho ^2-2 \rho  \bar{\rho}\right) \edth
  -4\left(6 \tau  \rho ^2-2 \bar{\rho} \tau  \rho -3 \bar{\rho}^2 \tau \right)
  \big]h_{nn}
  \nonumber \\ & 
  + \big[
  4 \edth'\edth\th
  +6 \th'\th\th
  +2(2\bar{\rho}'-17 \rho') \th\th
  +16(\rho -\bar{\rho}) \th'\th
  +4(\tau'- \bar{\tau}) \edth\th
  -24 \tau  \edth'\th
  +4(2 \rho - \bar{\rho}) \edth'\edth
  \nonumber \\ & \quad
  + \left(4\rho' \rho(8 \bar{\rho} -5 \rho) +8\tau \rho (2 \bar{\tau} -5 \tau')+4\bar{\rho}^2 (\rho'-2 \bar{\rho}')+24 \bar{\rho} \tau  \bar{\tau} -8\psi_2 (5 \rho+3 \bar{\rho}) +28 \bar{\psi}_2 \rho -8\bar{\rho} \bar{\psi}_2-20 \rho \bar{\psi}_2 \bar{\zeta} /\zeta \right)
  \nonumber \\ & \quad
  +4\left(5 \rho ^2-6 \bar{\rho} \rho + \bar{\rho}^2\right) \th'
  +16 \rho  \tau' \edth
  -8(5 \rho  \tau -2\bar{\rho} \tau + \bar{\rho} \bar{\tau}') \edth'
  \nonumber \\ & \quad
  +\left(24 \rho'(3 \bar{\rho}-2 \rho)-16 \bar{\rho} \bar{\rho}'+8\tau (3 \bar{\tau}-\tau')-16 \bar{\tau} \bar{\tau}'+36 \psi_2+8 \bar{\psi}_2-8\bar{\psi}_2 \bar{\zeta}/\zeta+8 \bar{\psi}_2 \bar{\zeta}^2/\zeta ^2\right) \th
  \big] h_{nm}
  \nonumber \\ & 
  + \big[
  -2 \edth\edth\th
  +4\left(5 \tau ^2-6 \bar{\tau}' \tau + \bar{\tau}'^2\right)\th
  +8(\rho  \tau - \bar{\rho} \tau + \bar{\rho} \bar{\tau}') \edth
  +4(2 \tau + \bar{\tau}') \edth\th
  +2( \bar{\rho}-2 \rho ) \edth\edth
  \nonumber \\ & \quad
  +12\left(4 \rho  \tau ^2+ \bar{\rho} \tau ^2-2 \bar{\rho} \bar{\tau}' \tau +\bar{\rho} \bar{\tau}'^2\right) 
  \big]h_{n\mb}
  \nonumber \\ & 
  + \big[
   \edth'\edth'\edth
  +6 \edth'\th'\th
  +12( \tau'-\bar{\tau}) \th'\th
  +8( \bar{\rho}'-4 \rho') \edth'\th
  +8( \rho - \bar{\rho}) \edth'\th'
  +2(2 \tau'- \bar{\tau}) \edth'\edth
  -6 \tau  \edth'\edth'
  \nonumber \\ & \quad
  +\left(36 \rho' \bar{\tau}(\bar{\rho}-2 \rho)-4\bar{\tau}(4 \tau  \bar{\tau}+3 \bar{\rho} \bar{\rho}')+8 \tau'(\tau \bar{\tau}-6 \rho  \rho' )-2\psi_2 (7 \bar{\tau}-4 \tau')+2 \bar{\tau}\bar{\psi}_2 +2\tau' \bar{\psi}_2(5 -9 \bar{\zeta}/\zeta)\right)
  \nonumber \\ & \quad 
  +2(8 \rho' \bar{\tau}-3 \bar{\rho}' \bar{\tau}-26 \rho' \tau') \th
  +2(4 \rho  \bar{\tau}+4\bar{\rho} \bar{\tau}+15 \rho  \tau') \th'
  +2\left(3 \tau'^2-2 \bar{\tau}\tau'\right) \edth
  \nonumber \\ & \quad
  +\left(18\rho'(2 \bar{\rho} - \rho)-12 \bar{\rho} \bar{\rho}'+4\bar{\tau}(3 \tau -2 \bar{\tau}')-14 \tau  \tau'+2 \psi_2+5 \bar{\psi}_2-5\bar{\psi}_2 \bar{\zeta}/\zeta+4 \bar{\psi}_2 \bar{\zeta}^2/\zeta ^2\right) \edth'
  \big]h_{mm}
  \nonumber \\ & 
  + \big[
  -8 \edth\th'\th
  -2 \edth'\edth\edth
  +12( \tau + \bar{\tau}')\th'\th
  +24 \rho' \edth\th
  +12( \bar{\rho}- \rho ) \edth\th'
  +2( \bar{\tau}- \tau') \edth\edth
  +4(2 \tau + \bar{\tau}') \edth'\edth
  \nonumber \\ & \quad
  +\left(20 (\bar{\tau} \bar{\tau}'^2+ \tau ^2 \tau')+40( \rho  \rho' \tau + \bar{\rho} \bar{\rho}' \bar{\tau}')+24 \bar{\rho} \tau (\bar{\rho}' - \rho') -28\tau  \bar{\tau} ( \tau+ \bar{\tau}')+8 \psi_2(7 \tau+3 \bar{\tau}')-4 \bar{\tau}' \bar{\psi}_2(1+3\bar{\zeta}^2/\zeta ^2)\right) 
  \nonumber \\ & \quad
  +4(9 \rho'\tau +10 \bar{\rho}' \tau +5 \bar{\rho}' \bar{\tau}') \th
  +4(3 \rho  \tau -7\bar{\rho} \tau +4 \bar{\rho} \bar{\tau}') \th'
  +4\left(5 \tau ^2-6 \bar{\tau}' \tau + \bar{\tau}'^2\right) \edth'
  \nonumber \\ & \quad
  +\left(4 \rho'(\rho  -3 \bar{\rho})+8 \bar{\rho} \bar{\rho}'-4 \tau (\bar{\tau}+  \tau')+12 \bar{\tau} \bar{\tau}'+6 \psi_2-4 \bar{\psi}_2+6 \bar{\psi}_2 \bar{\zeta}/\zeta-8 \bar{\psi}_2 \bar{\zeta}^2/\zeta^2\right) \edth
  \big] h_{m\mb}
  \nonumber \\ & 
  + \big[
  \edth\edth\edth
  -2\left(7 \tau ^2-4 \bar{\tau}' \tau + \bar{\tau}'^2\right) \edth
  -2( \tau + \bar{\tau}') \edth\edth
  -4\left(6 \tau ^3-5 \bar{\tau}' \tau ^2+ \bar{\tau}'^2 \tau \right)
  \big] h_{\mb\mb}
  \Big\} 
\end{align}%
\endgroup
The adjoint operator has four components given by symmetries:
\begin{align}
(\cA^\dag \xi)_{nn} &=- (\cA^\dag \xi)_{ll}',\\
(\cA^\dag \xi)_{\mb\mb} &=- (\cA^\dag \xi)_{mm}',\\
(\cA^\dag \xi)_{n\mb} &=- (\cA^\dag \xi)_{lm}',\\
(\cA^\dag \xi)_{nm} &=- (\cA^\dag \xi)_{l\mb}',
\end{align}
and a fifth component which is simply relate to another component
\begin{equation}
(\cA^\dag \xi)_{m\mb} = -(\cA^\dag \xi)_{ln}
 -\frac19 \psi_2\big[
  (2 \th' + \rho' - 2\bar{\rho}')\zeta^4 \xi_l
  -(2 \th + \rho - 2\bar{\rho})\zeta^4 \xi_n
  -(2 \edth' + \tau' - 2\bar{\tau})\zeta^4 \xi_m
  +(2 \edth + \tau - 2\bar{\tau}')\zeta^4 \xi_\mb
  \big].
\end{equation}
The other five components are given by
\begingroup
\allowdisplaybreaks
\begin{align}
 (\cA^\dag \xi)_{ll} &= \frac{\zeta^4}{108} \Big\{
  \big[
  -\th'\th\th
  -6 \edth'\edth\th
  +6(7\rho- \bar{\rho}) \edth'\edth
  +14 \rho \th'\th
  +(\bar{\rho}'-\rho') \th\th
  -2 (\tau-7 \bar{\tau}') \edth'\th
  -2 (\bar{\tau}-19 \tau') \edth\th
  \nonumber \\ & \quad
  +\left(12 \rho \rho'-26 \rho' \bar{\rho}+14 \bar{\rho}' \bar{\rho}-4 \tau \tau'-74 \tau \bar{\tau}+12 \bar{\tau}' \bar{\tau}+4 \psi_2+8 \bar{\psi}_2+ \bar{\psi}_2 \bar{\zeta}/\zeta-7 \bar{\psi}_2 \bar{\zeta}^2/\zeta^2\right)\th
  -42 \rho^2 \th'
  \nonumber \\ & \quad
  +(6 \tau (2 \rho+9 \bar{\rho})+8 \bar{\rho} \bar{\tau}') \edth'
  -2 (90 \rho \tau'+9 \rho \bar{\tau}+4 \bar{\rho} \bar{\tau}) \edth
  +\big( 6 \rho\rho' (14\bar{\rho}-5\rho)-2\bar{\rho}^2 (31 \rho'-4 \bar{\rho}')
  \nonumber \\ & \qquad
 +4\rho \tau (5\tau'+43 \bar{\tau})+12 \bar{\rho} \bar{\tau}(\tau +\bar{\tau}')-16 \rho \psi_2+27 \bar{\rho} \psi_2+4 \bar{\rho} \bar{\psi}_2-27 \rho \bar{\psi}_2-10 \rho \bar{\psi}_2\bar{\zeta} /\zeta\big)
  \big]\xi_l
  \nonumber \\ &
  + \big[
  \th\th\th
  -(13 \rho+\bar{\rho}) \th\th
  + 2\left(7 \rho \bar{\rho}+15 \rho^2- \bar{\rho}^2\right) \th
  +2\left(15 \rho^3-21 \rho^2 \bar{\rho}+7 \rho \bar{\rho}^2-\bar{\rho}^3\right)
  \big] \xi_n
  \nonumber \\ &
  + \big[
  7 \edth'\th\th
  -6 (9 \rho+\bar{\rho}) \edth'\th
  -(37 \tau'+\bar{\tau}) \th\th
  -2 (9 \rho \bar{\tau}+4 \bar{\rho} \bar{\tau}-86 \rho \tau') \th
  +6 \left(7 \rho \bar{\rho}+5 \rho^2-\bar{\rho}^2\right) \edth'
  \nonumber \\ & \quad
  +2 \left(15 \rho \bar{\rho}\bar{\tau} +48 \rho^2\bar{\tau} -7 \bar{\rho}^2\bar{\tau} +5 \rho^2 \tau'\right)
  \big] \xi_m
  \nonumber \\ &
  + \big[
  -\edth\th\th
  +2(7 \rho-\bar{\rho}) \edth\th
  +(3 \bar{\tau}'-\tau) \th\th
  +2(6 \rho \tau+13 \bar{\rho} \tau+2 \bar{\rho} \bar{\tau}') \th
  -2 \left(21 \rho^2-7 \rho \bar{\rho}+\bar{\rho}^2\right) \edth
  \nonumber \\ & \quad
  -2 \left(3 \tau \left(5 \rho \bar{\rho}+5 \rho^2-2 \bar{\rho}^2\right)-\bar{\rho}^2 \bar{\tau}'\right)
  \big] \xi_\mb
   \Big\},
\\
(\cA^\dag \xi)_{mm} &= \frac{\zeta^4}{108} \Big\{
  \big[
  -7 \edth\edth\th'
  +(23 \rho'+\bar{\rho}') \edth\edth
  +(68 \tau+6 \bar{\tau}') \edth\th'
  +4 (3 \bar{\rho}' \tau-26 \rho' \tau+2 \bar{\rho}' \bar{\tau}') \edth
  +6 \left(\bar{\tau}'^2-8 \tau \bar{\tau}'-14 \tau^2\right) \th'
  \nonumber \\ & \quad
  -2 \left(45 \bar{\rho}' \tau^2 +15 \bar{\rho}'\bar{\tau}'\tau +25\rho' \tau^2 - 7 \bar{\rho}' \bar{\tau}'^2 \right)
  \big]\xi_l
  \nonumber \\ &
  + \big[
  \edth\edth\th
  +(\rho-\bar{\rho}) \edth\edth
  -14 \tau \edth\th
  + 4 (3 \tau (\bar{\rho}-\rho)-\bar{\rho} \bar{\tau}') \edth
  +42 \tau^2 \th
  +\left(30 \tau^2 (\rho-\bar{\rho})+26 \bar{\rho} \tau \bar{\tau}'-6 \bar{\rho} \bar{\tau}'^2\right)
  \big] \xi_n
  \nonumber \\ &
  + \big[
  6 \edth\th'\th
  +\edth'\edth\edth
  -48 \tau \th'\th
  -14 \tau \edth'\edth
  +2 (\rho-4 \bar{\rho}) \edth\th'
  +2(\bar{\rho}'-16 \rho') \edth\th
  +(\tau'-\bar{\tau}) \edth\edth
  \nonumber \\ & \quad
  + \left(4\bar{\rho}(7 \rho' - \bar{\rho}')+6 \rho \rho'-4 \tau (3 \tau'-5 \bar{\tau})-8 \bar{\tau}' \bar{\tau}+4 \psi_2+5 \bar{\psi}_2-\bar{\psi}_2\bar{\zeta}/\zeta+4 \bar{\psi}_2\bar{\zeta}^2/\zeta^2\right) \edth
  \nonumber \\ & \quad
  +2 (31 \bar{\rho} \tau-7 \rho \tau-7 \bar{\rho} \bar{\tau}') \th'
  +14 \tau (12 \rho'-\bar{\rho}') \th
  +42 \tau^2 \edth'
  +\big(6\tau^2(5  \tau'-14 \bar{\tau})+2\bar{\tau}' \bar{\tau}(34 \tau -7 \bar{\tau}')
  \nonumber \\ & \qquad
  -\rho' \tau(22 \rho +164 \bar{\rho})-\bar{\rho} \bar{\rho}' (20 \tau+26 \bar{\tau}')-\psi_2 (16 \tau -27 \bar{\tau}')-3  \tau \bar{\psi}_2 (9-2\bar{\zeta}/\zeta)+7 \bar{\tau}' \bar{\psi}_2(1+\bar{\zeta}^2/\zeta^2)\big)
  \big] \xi_m
  \nonumber \\ &
  + \big[
  -\edth\edth\edth
  +(13 \tau+\bar{\tau}') \edth\edth
  +2 \left(\bar{\tau}'^2-7 \tau \bar{\tau}'-15 \tau^2\right) \edth
  +2 \left(\bar{\tau}'^3-7  \tau \bar{\tau}'^2+21 \tau^2 \bar{\tau}'-15 \tau^3\right)
  \big] \xi_\mb
   \Big\},
\\
(\cA^\dag \xi)_{lm} &= \frac{\zeta^4}{108} \Big\{
  \big[
  -4 \edth\th'\th
  -3 \edth'\edth\edth
  +4(7 \rho- \bar{\rho}) \edth\th'
  +2 (7 \rho'-\bar{\rho}') \edth\th
  +2(5 \tau+3 \bar{\tau}') \th'\th
  +4 (5 \tau+\bar{\tau}') \edth'\edth
  \nonumber \\ & \quad
  +(19 \tau'-4 \bar{\tau}) \edth\edth
  +\left(2 \bar{\rho} (2 \rho'-\bar{\rho}')-56 \rho \rho'-2 \bar{\tau} (4 \tau- \bar{\tau}')-92 \tau \tau'+2 \psi_2+3 \bar{\psi}_2-21 \bar{\psi}_2 \bar{\zeta}/\zeta\right)\edth
  \nonumber \\ & \quad
  +(8 \tau (3 \bar{\rho}-8 \rho)+2 \bar{\rho} \bar{\tau}') \th'
  +2 (6 \rho' \tau+12 \bar{\rho}' \tau+7 \bar{\rho}' \bar{\tau}') \th
  +\left(-28 \tau \bar{\tau}'+6 \tau^2+4 \bar{\tau}'^2\right) \edth'
  +\big(6 \bar{\rho}' \bar{\rho} (4 \tau+3 \bar{\tau}')
  \nonumber \\ & \quad
  -4 \rho' \tau (12 \rho+\bar{\rho})-42 \tau \bar{\tau} (\bar{\tau}'-2 \tau)+10 \tau^2 \tau'+12 \bar{\tau}'^2 \bar{\tau}-4 \psi_2(2\tau+3 \bar{\tau}')-\bar{\psi}_2(35 \tau -2 \bar{\tau}')+3 \tau \bar{\psi}_2\bar{\zeta}/\zeta\big)
  \big]\xi_l
  \nonumber \\ &
  + \big[
  \edth\th\th
  -6 \rho \edth\th
  -7 \tau \th\th
  +(6 \tau (6 \rho+\bar{\rho})-2 \bar{\rho} \bar{\tau}') \th
  -2 \left(3 \rho^2-4 \rho \bar{\rho}+\bar{\rho}^2\right) \edth
  +6 \rho \tau(5 \rho-6 \bar{\rho})-2 \bar{\rho}^2 (4\tau+3 \bar{\tau}')
  \big] \xi_n
  \nonumber \\ &
  + \big[
  3 \th'\th\th
  +4 \edth'\edth\th
  -28 \tau \edth'\th
  -4 (5 \rho+\bar{\rho}) \th'\th
  -2 (3 \rho+\bar{\rho}) \edth'\edth
  +(4 \bar{\rho}'-19 \rho') \th\th
  +2 (\bar{\tau}-9 \tau') \edth\th
  \nonumber \\ & \quad
  + \left(4 \rho' (23 \rho+3 \bar{\rho})-6 \bar{\rho}' \bar{\rho}+2 \bar{\tau}(5 \tau-2 \bar{\tau}')+84 \tau \tau'+2 \psi_2+\bar{\psi}_2+21  \bar{\psi}_2\bar{\zeta}/\zeta+2  \bar{\psi}_2 \bar{\zeta}^2/\zeta^2\right)\th
  \nonumber \\ & \quad
  +(4 \tau (9 \rho+5 \bar{\rho})-6 \bar{\rho} \bar{\tau}') \edth'
  -2( \bar{\tau} (5 \rho+4 \bar{\rho})+2 \rho \tau') \edth
  +\left(28 \rho \bar{\rho}-6 \rho^2-4 \bar{\rho}^2\right) \th'
  +\big(4 \rho \tau (5\tau'+2 \bar{\tau})
  \nonumber \\ & \qquad
  +\bar{\rho} \bar{\tau}(38\tau -28 \bar{\tau}')- 2\rho\rho'\left(42  \bar{\rho} +5 \rho\right)+2\bar{\rho}^2(23 \rho' -8 \bar{\rho}')-4 \psi_2(2 \rho+3 \bar{\rho})- \rho \bar{\psi}_2(35-9\bar{\zeta}/\zeta)\big)
  \big] \xi_m
  \nonumber \\ &
  + \big[
  -\edth\edth\th
  +(7 \rho-2 \bar{\rho}) \edth\edth
  +(6 \tau+2 \bar{\tau}') \edth\th
  +12 \tau (\bar{\rho}-3 \rho) \edth
  +2 \left(3 \tau^2-7 \tau \bar{\tau}'+\bar{\tau}'^2\right) \th
  \nonumber \\ & \quad
  +\left(4 \bar{\rho} \bar{\tau}'^2-6 \tau^2 (5 \rho+6 \bar{\rho})-2 \bar{\rho} \tau \bar{\tau}'\right)
  \big] \xi_\mb
   \Big\},
\\
(\cA^\dag \xi)_{l\mb} &= \frac{\zeta^4}{108} \Big\{
  \big[
  2 \edth'\th'\th
  +3 \edth'\edth'\edth
  -12 \rho \edth'\th'
  -14 \tau' \th'\th
  +2( \rho'- \bar{\rho}') \edth'\th
  +(2 \tau-5 \bar{\tau}') \edth'\edth'
  +2 (\bar{\tau}-19 \tau') \edth'\edth
  \nonumber \\ & \quad
  +\left(2 \rho'(9 \bar{\rho}-4 \rho)-10 \bar{\rho}' \bar{\rho}+2\tau (25 \bar{\tau}-4 \tau')-8 \bar{\tau}' \bar{\tau}-20 \psi_2-\bar{\psi}_2(5+2\bar{\zeta}/\zeta-5\bar{\zeta}^2/\zeta^2)\right)\edth'
  \nonumber \\ & \quad
  -4 (3 \rho' \tau'+3 \rho' \bar{\tau}+\bar{\rho}' \bar{\tau}) \th
  +56 \rho \tau' \th'
  +2 \left(51 \tau'^2-7 \tau' \bar{\tau}+\bar{\tau}^2\right) \edth
  \nonumber \\ & \quad
  + \big(40 \rho \rho' (\tau'+2\bar{\tau})-6 \bar{\rho}\bar{\tau}(3 \bar{\rho}'+2 \rho')+6\bar{\tau}^2(6 \tau - \bar{\tau}') -10\tau \tau'(9 \bar{\tau}+2 \tau')+4\psi_2 (5\bar{\tau} +8\tau')-9\bar{\psi}_2(\bar{\tau} -3 \tau')\big)
  \big]\xi_l
  \nonumber \\ &
  + \big[
  -5 \edth'\th\th
  +2 (20 \rho+\bar{\rho}) \edth'\th
  +(20 \tau'+3 \bar{\tau}) \th\th
  +(8 \bar{\rho} \bar{\tau}-70 \rho \tau'-18 \rho \bar{\tau}) \th
  +2 \left(\bar{\rho}^2-6 \rho \bar{\rho}-25 \rho^2\right) \edth'
  \nonumber \\ & \quad
  +\left(2 \bar{\tau} \left(8 \rho^2-16 \rho \bar{\rho}+5 \bar{\rho}^2\right)-50 \rho^2 \tau'\right)
  \big] \xi_n
  \nonumber \\ &
  + \big[
  -5 \edth'\edth'\th
  +(10 \rho+3 \bar{\rho}) \edth'\edth'
  +2(25 \tau'+ \bar{\tau}) \edth'\th
  -4(10 \rho \tau'-5 \rho \bar{\tau}-2 \bar{\rho} \bar{\tau}) \edth'
  +2 \left(\bar{\tau}^2-45 \tau'^2-7 \tau' \bar{\tau}\right) \th
  \nonumber \\ & \quad
  +\left(2 \bar{\tau}^2 (16 \rho+5 \bar{\rho})-20 \rho \tau'^2-14 \rho \tau' \bar{\tau}\right)
  \big] \xi_m
  \nonumber \\ &
  + \big[
  3 \th'\th\th
  +2 \edth'\edth\th
  +2 (\bar{\rho}-6 \rho) \edth'\edth
  -36 \rho \th'\th
  -3 \bar{\rho}' \th\th
  +2(4 \tau-5 \bar{\tau}') \edth'\th
  +(6 \bar{\tau}-20 \tau') \edth\th
  \nonumber \\ & \quad
  + \left(2 \rho' (\rho+19 \bar{\rho})-10 \bar{\rho}' \bar{\rho}-2 \tau(9 \tau'-20 \bar{\tau})-4 \bar{\tau}' \bar{\tau}+22 \psi_2- \bar{\psi}_2(8+\bar{\zeta}/\zeta- 5 \bar{\zeta}^2 /\zeta^2)\right)\th
  \nonumber \\ & \quad
  -8 (\bar{\rho} \bar{\tau}'+5 \tau (\rho+\bar{\rho})) \edth'
  +(86 \rho \tau'-18 \rho \bar{\tau}+8 \bar{\rho} \bar{\tau}) \edth
  +90 \rho^2 \th'
  + \big(8 \bar{\rho}^2(6\rho'- \bar{\rho}')-10 \rho\rho'(9 \bar{\rho}+ \rho)
  \nonumber \\ & \qquad
  +10 \rho \tau(6 \tau'-7 \bar{\tau})+4\bar{\rho} \bar{\tau}(2 \tau - \bar{\tau}')-10\psi_2(5 \rho +2 \bar{\rho})+ \rho \bar{\psi}_2(23+9 \bar{\zeta}/\zeta)-4 \bar{\rho} \bar{\psi}_2\big)
  \big] \xi_\mb
   \Big\},
\\
(\cA^\dag \xi)_{ln} &= \frac{\zeta^4}{108} \Big\{
  \big[
  \th'\th'\th
  +4 \edth'\edth\th'
  -6 \rho \th'\th'
  -6 \rho' \th'\th
  +2(\bar{\rho}'-7 \rho') \edth'\edth
  -4 (2 \tau+\bar{\tau}') \edth'\th'
  +2 (\bar{\tau}-13 \tau') \edth\th'
  \nonumber \\ & \quad
  +2 \left(2 \rho' (5 \rho+3 \bar{\rho})-2 \bar{\rho}' \bar{\rho}+23 \tau \tau'+8 \tau \bar{\tau}-\bar{\tau}' \bar{\tau}-\bar{\psi}_2(1-4 \bar{\zeta}/\zeta-\bar{\zeta}^2/\zeta^2)\right)\th'
  -2 \left(3 \rho'^2-4 \rho' \bar{\rho}'+\bar{\rho}'^2\right) \th
  \nonumber \\ & \quad
  -2 (6 \rho' \tau+7 \bar{\rho}' \tau+5 \bar{\rho}' \bar{\tau}') \edth'
  -\rho' (6 \bar{\tau}-74 \tau') \edth
  +\big(2 \rho'^2 (8 \rho-21 \bar{\rho})+2\bar{\rho}' \bar{\rho}(18 \rho'-5 \bar{\rho}') +2\bar{\tau}\bar{\rho}'(11 \tau  -8 \bar{\tau}')
  \nonumber \\ & \qquad
  +2\rho' \tau (9\tau'-32 \bar{\tau})- \psi_2(28 \rho'+33 \bar{\rho}')+3\rho' \bar{\psi}_2(1+5 \bar{\zeta}/\zeta) -5 \bar{\rho}' \bar{\psi}_2\big)
  \big]\xi_l
  \nonumber \\ &
  + \big[
  -\th'\th\th
  -4 \edth'\edth\th
  +2 (\rho+2 \bar{\rho}) \th'\th
  +2(7 \rho- \bar{\rho}) \edth'\edth
  +2(5 \rho'-2 \bar{\rho}') \th\th
  +24 \tau \edth'\th
  +2 (5 \tau'+\bar{\tau}) \edth\th
  \nonumber \\ & \quad
  +\left(4 \rho' \bar{\rho}+2 \bar{\rho}' \bar{\rho}-34 \rho \rho'-44 \tau \tau'-18 \tau \bar{\tau}+4 \bar{\tau}' \bar{\tau}-\psi_2+\bar{\psi}_2(1-8 \bar{\zeta}/\zeta-2\bar{\zeta}^2/\zeta^2)\right)\th
  \nonumber \\ & \quad
  -2 (\tau (35 \rho+\bar{\rho})-\bar{\rho} \bar{\tau}') \edth'
  +(8 \rho \tau'-10 \rho \bar{\tau}+8 \bar{\rho} \bar{\tau}) \edth
  +4 \left(5 \rho^2-6 \rho \bar{\rho}+\bar{\rho}^2\right) \th'
  +\big(2\rho \rho' (22\bar{\rho}-15 \rho)
  \nonumber \\ & \qquad
  -2\bar{\rho}^2(11 \rho'-4 \bar{\rho}')- \rho \tau(20 \tau'-62 \bar{\tau})-2 \bar{\rho}\bar{\tau}(9 \tau -10 \bar{\tau}')+\psi_2(17 \rho +35 \bar{\rho})+\rho \bar{\psi}_2(17 -15 \bar{\zeta}/\zeta)\big)
  \big] \xi_n
  \nonumber \\ &
  + \big[
  -4 \edth'\th'\th
  -\edth'\edth'\edth
  +4 (\rho+\bar{\rho}) \edth'\th'
  +(26 \rho'-6 \bar{\rho}') \edth'\th
  +6 \tau \edth'\edth'
  +2 (9 \tau'+\bar{\tau}) \th'\th
  +6 \tau' \edth'\edth
  \nonumber \\ & \quad
  +\left(6 \bar{\rho}\bar{\rho}'-24 \rho \rho'-14 \bar{\rho}\rho' -24 \tau \tau'-8 \tau \bar{\tau}+4 \bar{\tau}\bar{\tau}' -22 \psi_2-2 \bar{\psi}_2(1+3 \bar{\zeta}/\zeta+ \bar{\zeta}^2/\zeta^2)\right)\edth'
  \nonumber \\ & \quad
  +2(4 \rho \tau'+5 \rho \bar{\tau}+3 \bar{\rho} \bar{\tau}) \th'
  -12 \rho' (8 \tau'+\bar{\tau}) \th
  +2 \left(3 \tau'^2-4 \tau' \bar{\tau}+\bar{\tau}^2\right) \edth
  +\big( 20\bar{\tau} \bar{\rho}' \bar{\rho}-2\bar{\tau} \rho' (29 \rho+9 \bar{\rho})
  \nonumber \\ & \qquad
  -2\bar{\tau}^2(18 \tau -5 \bar{\tau}')-2\tau \tau'(10 \tau'-23 \bar{\tau})-\psi_2(9 \bar{\tau}-58 \tau')-5\bar{\tau} \bar{\psi}_2+3\tau'\bar{\psi}_2 (1 - \bar{\zeta} /\zeta)\big)
  \big] \xi_m
  \nonumber \\ &
  + \big[
  4 \edth\th'\th
  +\edth'\edth\edth
  +4 (\bar{\rho}-6 \rho) \edth\th'
  -2 (3 \rho'+\bar{\rho}') \edth\th
  -2 (9 \tau+\bar{\tau}') \th'\th
  -2(\tau+2\bar{\tau}') \edth'\edth
  +2(2 \bar{\tau}-5 \tau') \edth\edth
  \nonumber \\ & \quad
  +\left(4 \rho' (5 \rho+3 \bar{\rho})-2 \bar{\rho}' \bar{\rho}+34 \tau \tau'-8 \tau \bar{\tau}+2 \bar{\tau}' \bar{\tau}+23 \psi_2-\bar{\psi}_2+8 \bar{\psi}_2\bar{\zeta}/\zeta\right)\edth
  \nonumber \\ & \quad
  +(94 \rho \tau-6 \bar{\rho} \tau+2 \bar{\rho} \bar{\tau}') \th'
  +(2 \bar{\rho}' (2 \tau-5 \bar{\tau}')-8 \rho' \tau) \th
  -4 \left(5 \tau^2-6 \tau \bar{\tau}'+\bar{\tau}'^2\right) \edth'
  +\big(2 \tau^2(15 \tau'-22 \bar{\tau})
  \nonumber \\ & \qquad
  +2\bar{\tau}' \bar{\tau}(9 \tau -2 \bar{\tau}')-2\rho' \tau(2 \rho +29 \bar{\rho})-2 \bar{\rho}' \bar{\rho} (6 \tau+7 \bar{\tau}')-\psi_2(67 \tau -11 \bar{\tau}')+\tau \bar{\psi}_2(17 +7 \bar{\zeta}/\zeta)-2 \bar{\tau}' \bar{\psi}_2\big)
  \big] \xi_\mb
   \Big\}.
\end{align}%
\endgroup
\end{subequations}

\subsection{\texorpdfstring{$\cB_0$}{B0} and \texorpdfstring{$\cB_4$}{B4} operators}

The operators $\cB_0$ and $\cB_4$ appearing in Eq.~\eqref{eq:TN} are given by
\begin{subequations}
\begin{align}
  \cB_0 \xi &= \frac19 \zeta^4 \big[\tau \edth \th - \rho \edth^2-2 \tau \bar{\tau}' \th + 2 \rho \bar{\tau}'^2\big]\xi_l
  + \frac19 \zeta^4 \big[ \rho \th \edth  - \tau \th^2 - 2 \rho \bar{\rho} \edth + 2 \tau \bar{\rho}^2 \big] \xi_m, \\
  \cB_4 \xi &= -\frac19 \zeta^4 \big[\tau' \edth' \th' - \rho' \edth'^2-2 \tau' \bar{\tau} \th' + 2 \rho' \bar{\tau}^2\big]\xi_n
  - \frac19 \zeta^4 \big[ \rho' \th' \edth'  - \tau' \th'^2 - 2 \rho' \bar{\rho}' \edth' + 2 \tau' \bar{\rho}'^2 \big] \xi_\mb = -\cB_0'.
\end{align}
Their adjoints, appearing in Eq.~\eqref{eq:NT}, are given by
\begin{align}
  (\cB_0^\dag \Psi_4)^\alpha &= \frac19 \zeta^4 l^\alpha\big[\tau \th \edth - \rho \edth^2+\tau(\bar{\tau}' -4\tau) \th+4\rho \tau \edth\big] \Psi_4
  + \frac19 \zeta^4 m^\alpha\big[ \rho \edth \th  - \tau \th^2 +\rho(\bar{\rho} -4\rho) \edth+4\rho \tau \th\big] \Psi_4, \\
  (\cB_4^\dag \Psi_0)^\alpha &= - \big[(\cB_0^\dag \Psi_4)'\big]^\alpha \nonumber \\
  & \hspace{-1em} -\frac19 \zeta^4 n^\alpha\big[\tau' \th' \edth' - \rho' \edth'^2+\tau'(\bar{\tau} -4\tau') \th'+4\rho' \tau' \edth'\big] \Psi_0
  - \frac19 \zeta^4 \mb^\alpha\big[ \rho' \edth' \th'  - \tau' \th'^2 +\rho'(\bar{\rho}' -4\rho') \edth'+4\rho' \tau' \th'\big] \Psi_0.
\end{align}
\end{subequations}

\end{widetext}

\bibliography{DirectLorenzGauge}
\end{document}